\documentstyle[amssymb,prc,aps,epsfig]{revtex}

\begin{document}
\title{Correlation search for coherent pion emission in heavy ion collisions}
\author{S.V. Akkelin$^{2}$, R. Lednicky$^{1,3}$ and Yu.M. Sinyukov$^{1,2}$}
\maketitle

\begin{abstract}
The methods allowing to extract the coherent component of pion emission
conditioned by the formation of a quasi-classical pion source in heavy ion
collisions are suggested. They exploit a nontrivial modification of the
quantum statistical and final state interaction effects on the correlation
functions of like and unlike pions in the presence of the coherent
radiation. The extraction of the coherent pion spectrum from $\pi ^{+}\pi
^{-}$ and $\pi ^{\pm }\pi ^{\pm }$ correlation functions and single--pion
spectra is discussed in detail for large expanding systems produced in
ultra-relativistic heavy ion collisions.
\end{abstract}

\begin{center}
{\small {\it $^{1}$ SUBATECH, (UMR, Universite, Ecole des Mines,
IN2P3/CNRS), \\[0pt]
4, rue Alfred Castler, F-44070 Nantes Cedex 03, France. \\[0pt]
$^{2}$ Bogolyubov Institute for Theoretical Physics, Kiev 03143,
Metrologicheskaya 14b,Ukraine.\\[0pt]
$^{3}$ Institute of Physics ASCR, Na Slovance 2, 18221 Prague 8, Czech
Republic.\\[0pt]
}}
\end{center}

\section{Introduction}

The hadronic observables, such as single- or multi-particle hadron spectra,
play an important role in the studies of ultra-relativistic heavy ion
collisions. However, these observables contain rather indirect information
on the initial stage of the collision process since the particle interactions
result in substantial stochastization and thermalization
of a system during its evolution. Nevertheless, the final hadronic state
can carry some residual signals of the earlier stages of the particle
production process. A partial coherence of the produced pions is supposed to
be one of the important examples.

The first systematic study of coherent processes in high energy
hadron-nucleus ($h+A$) collisions was based on Glauber theory \cite{Glauber}.
In this theory, the $h+A$ collision is considered as a process of
subsequent scatterings of the projectile on separate nucleons of the
nucleus; the projectile energies are supposed much higher than the inverse
nucleus radius ($E_{h}\gg 1/R$), thus allowing to consider a linear projectile
trajectory inside the nucleus (eikonal approximation). If the scattering
process occurred with almost no recoil of the nucleus nucleons, i.e. with no
{\it witnesses} of the individual scatterings, then the $h+A$
collision should be described by a coherent superposition of the elementary
hadron-nucleon scattering amplitudes. Such a type of the collision is called
coherent scattering. Since the nucleus in coherent scattering
does not change its state, it manifests itself just as a particle with some
form-factor. In the oscillator approximation, the nucleus form-factor
can be represented by a Gaussian: $\exp (-{\bf q}^{2}R^{2}/4)$.
The coherent processes are essential only for small momenta transferred from
the projectile hadron to the nucleus: $\left| {\bf q}\right| <1/R.$
Then, one can neglect the recoil energy and consider the nucleus as a whole
during the scattering process. There is a kinematic limitation of the
minimal longitudinal momentum transfer, $\left| q_{z}\right| _{\min }\approx
(M^{2}-m_{h}^{2})/(2\left| {\bf p}_{h}\right| )$, required to produce a
particle or a group of particles of the invariant mass $M$. The vanishing of
$\left| q_{z}\right| _{\min }$ with the increasing energy explains why the
coherent processes can take place only at high enough energies. It is worth
noting that the total coherent cross-section does not die out with the
increasing energy (see., e.g., \cite{Nikitin}).\footnote{%
We are grateful to V. L. Lyuboshitz for drawing our attention to this
important point and for an interesting discussion.}

Typically, however, the transferred momenta are sufficient for substantial
{\it recoil} effects and the excitation of the nucleus or its breakup.
Then, due to a small
{\it coherence length} $\sim 1/\left| {\bf q}\right|$,
the nucleus does not participate in the collision as a whole and one can
consider the $h+A$ collision as an incoherent superposition of
elementary hadron-nucleon scatterings corresponding to random phases of the
amplitudes of the latter. The resulting cross-section is then given by the
sum of the moduli squared of the amplitudes (probabilities) at each of the
possible scattering points (unlike to coherent scattering, when the
individual amplitudes are summed up first). As a result, the particles are
produced in chaotic (incoherent) states.

Let us come back to the production of particles (e.g., pions) in the
processes of non-elastic coherent scattering at small transferred momenta.
Since the nucleus is not excited in these processes and manifests itself as
a quasi-classical object, one can describe particle production using the
quantum field model of interaction with a classical source \cite{thirring}.
It is well known that the interaction with a classical source results in the
production of bosons in coherent states \cite{Carruthers}. These states
minimize the uncertainty relation and, so, are the closest to
classical ones.\footnote{%
The {\it coherent states} have been introduced and studied in detail by
Glauber \cite{Glauber-1}. The concept of coherent states was then applied to
pion production in high energy processes in Refs. \cite{Horn,Botke,GKW}.}
This is the main physical link between the processes of coherent scattering
and particle production in coherent states.

In heavy ion collisions at high energies, the average multiplicities are
quite high, e.g., several thousands of pions can be produced at maximal RHIC
energies. The inclusive particle spectra thus represent natural
characteristics of these processes. A convenient way to account for the
coherent properties of these processes consists in a model description of
particle emission, rather than in detailed evaluation of the contributing
amplitudes. The Gyulassy-Kauffmann-Wilson (GKW) model \cite{GKW} is an
example of such an approach. The model assumes that all pions are radiated
by classical currents (sources) which are produced in some space--time
region during the collision process. The corresponding density
matrix is constructed by averaging over the unobservable positions
of the centers of individual sources.
The pion spectra then effectively contain both chaotic and
coherent components.
In fact, the chaotic component dominates in case of a large
emission region, while, in the opposite limit of very small space--time
extent of this region, almost all pions are produced in the coherent state.
This seems to be rather general result: if the distances between the centers
of pion sources are smaller than the typical wave length of the quanta (the
source size), the substantial overlap of the wave packets leads to the
strong correlations (indistinguishability) between the phases in pion wave
functions and, thus, to the coherence\cite{Sinyukov,llp}.

Recently, the coherence of multipion radiation in high energy heavy ion
collisions was studied within GKW model in Ref. \cite{Ryskin}. In the model,
due to the longitudinal Lorentz contraction of the colliding nuclei, almost
all pions produced with small transverse momenta $p_{t}<1/R$ in central
nucleus-nucleus collisions are emitted coherently, and their momentum
spectra are determined by the system's space--time extent.
Clearly, the coherence of pions can be destroyed by pion rescatterings.
Nevertheless, the duration of hadron formation may happen to be
long enough to allow a considerable part of the coherent pions escape from
the interaction zone without rescatterings \cite{Ryskin}. However, as noted
in \cite{Ryskin}, one can expect a strong suppression of the GKW mechanism
of coherent pion production if quark-gluon plasma were created: the
hadronization then occurs in a thermal quark-gluon system and hadrons are
produced in the chaotic state only. Note that clear signals of the
thermalization and collective flows, observed at CERN SPS and RHIC energies
(see, e.g., \cite{QM99,QM} and references therein), point to strong
rescattering effects and may reflect also the importance of the quark-gluon
degrees of freedom.

The new physical phenomena, expected in RHIC and LHC experiments
with heavy ions, are associated with the creation of quasi-macroscopic, very
dense and hot systems. In such systems, the deconfinement phase transition
and the restoration of the chiral symmetry are likely to happen, possibly
leading to creation of the new states of matter: quark-gluon plasma (QGP)
and disoriented chiral condensate (DCC). In the latter case, another
possibility for the coherent pion radiation (above the thermal background)
appears. If the DCC were created at the chiral phase transition, a
quasi--classical pion field $\stackrel{\rightarrow }{\pi }_{cl}$ forms the
ground state of the system. The subsequent system decay is accompanied by a
relaxation of the ground state to normal vacuum. Such a process can be
described by the quantum field model of interaction with a classical source
(see, e.g. \cite{Bjorken}), and results in the coherent pion radiation. One
of the general conditions of the ground state rearrangement and formation of
the quasi--classical field is a large enough system volume \cite{Migdal}.
Therefore, such a field could be generated in heavy ion collisions at
sufficiently high energies provided the spontaneous chiral symmetry breaking
via DCC formation takes place. The overpopulation of the (quasi) pion
medium, making it close to the Bose-Einstein condensation point, can lead to
the strengthening of the coherent component conditioned by the ground state
(quasi-particle vacuum) decay \cite{Akkelin}. Since the DCC appears
relatively late (at the end of the hadronization stage), the coherent
radiation could partially survive and be observed.

The coherent emission manifests itself in a most direct way in the
inclusive correlation function $C(p,q)$ of two
identical bosons in the region of very small $|{\bf q}|$; $%
p=(p_{1}+p_{2})/2$, $q=p_{1}-p_{2}$. In case of only chaotic contribution,
the intercept of the quantum statistical (QS) Bose-Einstein part of the
correlation function
$C_{QS}(p,0)=2$ \cite{GGLP} while, in the presence of the coherent radiation,
$C_{QS}(p,0)<2$. Generally, the coherence means strong phase correlations of
different radiation components. In Ref. \cite{Sinyukov}, a simple
quantum--mechanical model of the phase--correlated one-particle wave
packets with different radiation centers has been considered.
In such a case (corresponding to indistinguishable correlated
emitting centers), the emission amplitude $A(p)$ averaged over the
event ensemble is not equal to zero, $\langle A(p)\rangle \neq 0$, and the
QS correlation function intercept $C_{QS}(p,0)<2.$ In the second
quantization representation (more adequate for processes of multi--boson
production), the analogous results take place for inclusive averages of the
quantum field operators: $\langle a(p)\rangle \neq 0$, $C_{QS}(p,0)<2$,
provided the radiation has a non-zero {\it coherent state} component. The
latter represents a superposition of the states of all possible boson
numbers at fixed phase relations.

In practice, most of the correlation measurements is done with
{\it charged} particles. However, charged bosons
cannot form the usual coherent state since it obviously violates the
super-selection rule. To overcome this difficulty, the generalized concept
of charge-constrained coherent states should be used
\cite{Botke,GKW,Bhaumik}.
Nevertheless, the correlations of charged bosons are usually
described with the help of ordinary (not charge-constrained)
coherent states \cite{Weiner-1,Heinz}
(see, however, Refs. \cite{Andreev,Nakamura}).
Our treatment of
two-pion correlations takes into account the
restrictions imposed by the super-selection rule and is based on the
density matrix formalism.

The density matrix approach gives the possibility to describe,
in a natural way, the chaotic radiation
(the initial state then corresponding to a local-equilibrium statistical
operator of quasi--particle excitations) and coherent emission
(arising due to the interaction with a classical source).
This approach can easily incorporate also the squeeze-state component of pion
radiation \cite{Weiner}, appearing due to the modification of the
energy spectrum
of quasi-pions as compared with that of free pions \cite{Csorgo}.
The density matrix formalism is also simply related with the Wigner function
description of the multiparticle phase-space and its evolution governed by the
relativistic transport equation \cite{Groot}, representing very
useful tools with a clear classical limit. Recent development of the
classical current approach to multiparticle production
\cite{Weiner,Weiner-1} has made it closer to the
density matrix formalism; particularly,
the clasical current in momentum space has been shown
mathematically identical with the coherent-state representation of the
density matrix, the latter called ''$P$'' or Glauber-Sudarshan representation
\cite{Glauber-1}, see also \cite{Zhang}.

In our approach, the super-selection rule requires an averaging, in the
density matrix, over all orientations of the quasi-classical pion source in
the isospin space. As a consequence, the averaged pion field vanishes: $%
\langle a(p)\rangle =0$ whereas, for identical pions, the intercept $%
C_{QS}(p,0)$ is still less than 2. The correlations of non-identical pions
also appear to be sensitive to the presence of the quasi--classical source.
This sensitivity arises due to properties of the generalized coherent states
satisfying, after the averaging over all orientations of the quasi-classical
source in isospin space,
the super-selection rule for charged particles.
Due to isospin symmetry of the strong-interaction Hamiltonian, there
are unique relations for the intercepts $C_{QS}^{ij}(p,0)$
of the pure QS correlation functions of two pions in various charge states
$i,j={\pm},{0}$.
For example, the
coherence suppression of $C^{\pm \pm }$ determines the coherence enhancement
of $C^{+-}$.

The coherence phenomena can be, however, masked by a number of effects
suppressing the measured correlation functions. The most important among
them are the decays of long-lived particles and resonances ({e.g.}, $\Lambda
$, $K_{s}^{0}$, $\eta $, $\eta ^{\prime }$, $\dots $), the single- and
two--track resolution and particle contamination. In Ref.\cite{Heinz-1}, the
method to discriminate between the effects of coherent radiation and decays of
long-lived resonances has been proposed.
The method assumes the simultaneous analysis of two- and three--particle
correlation functions of identical pions. The practical utilization of the
method is however difficult due to a low statistics of near--threshold
three-pion combinations and the problem of the three--particle Coulomb
interaction; also, one has to account for the super--selection rule.\footnote
{The latter problems are absent for neutral pions. However, sufficiently
accurate measurements of neutral pion correlations are practically out of the
present experimental possibilities.
}
Therefore, in the present
work we will restrict ourselves to the consideration of two-particle
correlation functions.

In addition to QS, the correlations of particles with small
relative velocities are also influenced by their
final state interaction (FSI). The effect of the latter
on two--particle correlations
is well understood and introduces no principle problems. It is
important that the correlations in different two--pion systems are
influenced by the QS, FSI and coherence effects in a different way. This
offers a possibility to discriminate different effects suppressing the
measured correlation functions and so to extract the coherent contribution
using correlation functions of like and unlike pions measured at small
relative momenta.

In the paper we study the influence of the coherent pion radiation on the
behavior of pion inclusive spectra and two--pion correlation functions and,
based on it, develop the methods for the extraction of the coherent component
above the chaotic background. Despite we associate the coherent radiation
with the formation of the DCC (as the most probable mechanism of the
coherence in ultra--relativistic A+A collisions), our results are rather
general. Actually, they are based on the general properties of the coherent
pion radiation: the quasi-classical nature of the coherent pion source and
the constrains imposed by the charge super--selection rule.

In Sec. II, we consider a general form of the density matrix of partially
coherent pions, and calculate quantum statistical correlations of identical
and nonidentical pions.

In Sec. III, we set forth the density matrix formalism taking into account
the decays of short-lived resonances and FSI of produced pions, and
calculate the corresponding correlation functions.

In Sec. IV, we discuss how to extract the coherent component of particle
radiation from the two--pion correlation functions, particularly, in the
case of large expanding systems produced in ultra-relativistic A+A
collisions.

A short summary and conclusion are given in Sec. V.

\section{Quantum statistical correlations of partially coherent pions}

It is well known that the description of the inclusive pion spectra and
two-pion correlations is based on a computation of the following averages
\cite{GKW}:

\begin{equation}
\begin{array}{c}
\omega _{{\bf p}}\frac{d^{3}N_{i}}{d^{3}{\bf p}}\equiv
n_{i}(p)=\sum\limits_{\alpha }|{\cal T}(in;p,\alpha )|^{2}=\bigl\langle
a_{i}^{\dagger }({p})a_{i}({p)}\bigr\rangle , \\
\text{ } \\
\;\omega _{{\bf p}_{1}}\omega _{{\bf p}_{2}}\frac{d^{6}N_{ij}}{d^{3}{\bf p}%
_{1}d^{3}{\bf p}_{2}}\equiv n_{ij}(p_{1},p_{2})=\sum\limits_{\alpha }|{\cal T%
}(in;p_{1},p_{2},\alpha )|^{2}=\bigl\langle a_{i}^{\dagger }({p}%
_{1})a_{j}^{\dagger }({p}_{2})a_{i}({p}_{1})a_{j}({p}_{2})\bigr\rangle ,\text{ }
\\
\\
C^{ij}(p,q)=n_{ij}(p_{1},p_{2})/n_{i}(p_{1})n_{j}(p_{2}),\text{ }\omega _{%
{\bf p}_{i}}=\sqrt{m^{2}+{\bf p}_{i}^{2}},
\end{array}
\label{1}
\end{equation}
where ${\cal T}(in;p,\alpha )$ is the normalized invariant production
amplitude. The summation is done over all quantum numbers $\alpha $ of other
produced particles, including integration over their momenta; $%
a_{i}^{\dagger }({p})$ and $a_{i}({p)}$ are respectively the creation and
annihilation operators of asymptotically free pions $i={\pm},{0}$;
the bracket $\langle \dots \rangle $ formally corresponds to the averaging
over some density matrix $|f\rangle \langle f|.$ A special attention
requires the production of particles with near-by velocities which can be
strongly influenced particle interaction in the final state. In this
Section, we concentrate mainly on quantum statistical correlations
ignoring, for a while, the effects of resonance decays and FSI.

Let us suppose that the density matrix ${\bf \rho }$ is a statistical
operator describing the thermal hadronic system in a pre-decaying state on a
hyper-surface of the thermal freeze-out $\sigma _{f}:t=t_{f}({\bf x})$. After
the thermal freeze-out the system is out of local thermal equilibrium but
still can be in a pre-decaying (interacting) state. In fact, the complete
decay (neglecting the long-time scale forces) happens at some finite {\it %
asymptotic} times $t_{out}<\infty $. Then the formal solution of the Heisenberg
equation for the pionic annihilation (creation) operators at this post thermal
freeze-out stage has the form\footnote{%
For a space--like hypersurface $\sigma _{f}$ (an example is
$\sigma _{f}=t_{f}({\bf x})=(\tau^{2}+x_{long}^{2})^{1/2}$ in the Bjorken
hydrodynamic model with the proper expansion time $\tau $), the use of
the covariant Tomonaga--Schwinger formalism gives the same result with the
substitution $t\rightarrow t({\bf x})$.}:

\begin{equation}
{\rm a}_{i,qm}({\bf p},t_{out})=[{\rm a}_{i,qm}({\bf p}{,}t_{f})+{\rm d}_{i}(%
{\bf p,}t_{f},t_{out})]e^{-i\omega _{{\bf p}}(t_{out}-t_{f})}.  \label{1.1}
\end{equation}
It formally corresponds to the sum of the general solution of the free
(homogeneous) Heisenberg equation of motion for pionic field (first term),
and a particular solution of the complete (inhomogeneous) Heisenberg equation
with a source (second term). The value ${\rm d}_{i}({\bf p,}%
t_{f},t_{out})$ depends on the actual form of the source term in the Heisenberg
equation.

The decay of the system at this stage, $t_{f}<t<t_{out}$, can be accompanied by
the coherent pion radiation due to the modification of hadron
properties in hot and dense hadronic environment or - due to some
peculiarities of the phase transition from QGP to hadron gas, e.g., the
formation of DCC. In both cases, almost non--interacting quasiparticle
excitations could be formed above a rearranged ground state (''condensate'').

In the systems
containing the DCC, the appearance of the quasi-classical pion field $%
\stackrel{\rightarrow }{\pi }_{cl}$  (corresponding to the density of
virtual pionic excitations of the quasi-pionic vacuum) at the thermal stage is
usually described in the mean field approximation as $\pi _{i,cl}(x)=\pi
_{i}(x)-\pi _{i,qm}(x),$ where the field $\pi _{i,qm}(x)$ corresponds to the
quasi-pion quantum excitations above the temporary vacuum background
$\pi _{i,cl}(x)$ (the order parameter).
Assuming the isotopic symmetry of the Lagrangian like in the sigma model
(see, e.g., \cite{Randrup}), we have $\pi _{i,cl}(x)=e_{i}\pi _{cl}(x)$,
where ${\bf e}$ is randomly oriented unit vector, ${\bf e}^{2}=1$, in the
three-dimensional isospin space. Then, for each ${\bf e-}$orientation of the
quasi-pionic vacuum at the thermal freeze--out, the free quasi--pions $%
\stackrel{\rightarrow }{\pi }_{qm}$are distributed according to the Gibbs
local-equilibrium density matrix $\rho _{{\bf e}}$ above the quasi-pionic
vacuum. After the thermal freeze-out, when the decay of such a thermal
system happens, the quasi-pion masses approach
the usual free particle values
and the condensate (the temporary {\it disoriented} vacuum) tends to
relax back to the normal vacuum by emitting physical pions in coherent
states - the vacuum for quasi-particles becomes a coherent state for free
particles. The latter process is similar to particle radiation by a
classical source.

Then the ''source'' term in Eq. (\ref{1.1}) takes on the form
\begin{equation}
{\rm d}_{i}({\bf p,}t_{f},t_{out})={\rm d}_{i,qm}({\bf p,}t_{f},t_{out})
+e_{i}{\rm d}_{coh}({\bf p,}t_{f},t_{out}),\quad e_{0}=\cos
\theta \text{ , }e_{\pm }=\frac{\sin \theta }{\sqrt{2}}e^{\pm i\phi },
\label{1.2}
\end{equation}
where ${\rm d}_{i,qm}({\bf p,}t_{f},t_{out})$ and $e_{i}{\rm d}_{coh}({\bf p,%
}t_{f},t_{out})$ are q- and c-value quantities respectively.
While the total number of
pions of momentum ${\bf p}$ radiated by a classical source is
fixed by $\left| {\rm d}_{coh}({\bf p,}t_{f},t_{out})\right| ^{2}$, the
distribution of radiating pions in isospace is determined by the
orientation of the vector ${\bf e}$;
we suppose ${\bf e}$ independent of $x$.
We further assume that the quasi-pion masses at the thermal
freeze-out are near the physical mass, $m_{i}(t_{f})\simeq m_{out}\equiv m$,
neglecting a possible mass shift which can generate squeeze-state components
in particle radiation.\footnote{%
Squeeze-state component can arise also in a strongly inhomogeneous thermal
boson system for particles with wavelengths larger than the system's
homogeneity lengths \cite{Sinyu}. Below we will assume the pion Compton
wave--length much smaller than the typical system lengths of homogeneity
(e.g., hydrodynamical lengths) at the thermal freeze-out hypersurface $%
\sigma _{f}$.}
We will neglect the rescatterings at the post thermal
freeze-out stage,
i.e. put ${\rm d}_{i,qm}({\bf p,}t_{f},t_{out})\approx 0$,
and approximately describe
the production of coherent pions at this stage by
the quantum field model of the interaction
with a classical source \cite{thirring}. Then, there is well known
linear relationship between the annihilation (creation) operators
diagonalizing the pion field Hamiltonian at the times $t_{f\text{ }}$ and $%
t_{out}$ ($i={\pm },{0}$):
\begin{equation}
{\rm a}_{i,qm}({\bf p},t_{out})=[{\rm a}_{i,qm}({\bf p}{,}t_{f})+e_{i}{\rm d}%
_{coh}({\bf p,}t_{f},t_{out})]e^{-i\omega _{{\bf p}}(t_{out}-t_{f})},\quad
\label{2}
\end{equation}
where the c-value quantity ${\rm d}_{coh}({\bf p,}t_{f},t_{out})$ depends on a
mechanism and the rate of the classical field decay.\footnote{%
It follows, from the continuity of the complete field $\pi _{i}(x)$ and its
derivative at $t=t_{f}$ that, for a fast freeze-out ($%
t_{out}-t_f\rightarrow 0$), the quantity ${\rm d}_{coh}({\bf p,}%
t_{f},t_{out}) $ is directly associated with the strength of the pion
condensate. On the other hand, an adiabatically slow switch-off of the
classical source yields
${\rm d}_{coh}({\bf p,}t_{f},t_{out})\approx 0$
\cite{thirring}.}

The operators $a_{i}(p)$ of the asymptotic free pion field (with the origin
of the time coordinate shifted to the point $t_{f}$) are connected with the
operators ${\rm a}_{i,qm}({\bf p},t)$ taken at the {\it asymptotic} times $%
t_{out}$ by the relation \cite{Bogolubov}
\begin{equation}
a_{i}(p)=\sqrt{p_{0}}e^{ip_{0}(t_{out}-t_{f})}{\rm a}_{i,qm}({\bf p}%
,t_{out}),~~~p_{0}=\omega _{{\bf p}}.  \label{3}
\end{equation}
Eqs. (\ref{2})and (\ref{3}) allow to calculate the mean values of the
asymptotic operators $a_{i}(p)$ and $a_{i}^{\dagger }(p)$ for each ${\bf e}$%
{\bf -}orientation of the quasi-pion vacuum applying the thermal Wick theorem
to the operators ${\rm a}_{i,qm}({\bf p},t_{f})$ and ${\rm a}%
_{i,qm}^{\dagger }({\bf p},t_{f})$. The Gaussian form of the statistical
operator ${\bf \rho }_{{\bf e}}$ guarantees that $\langle {\rm a}_{i,qm}(%
{\bf p},t_{f})\rangle _{{\bf e}}=0$ for any fixed isospin orientation ${\bf %
e}$ of the quasi-particle vacuum. Then,
\begin{equation}
\begin{array}{c}
\bigl\langle a_{i}^{\dagger }({p}_{1})a_{j}^{\dagger }({p}_{2})a_{i}({p}%
_{1})a_{j}({p}_{2})\bigr\rangle _{{\bf e}}=
\bigl\langle a_{i}^{\dagger }({p}_{1})a_{i}({p}_{1})\bigr\rangle _{{\bf e}}
\bigl\langle a_{j}^{\dagger }({p}_{2})a_{j}({p}_{2})\bigr\rangle _{{\bf e}}+ \\
\\
\delta_{ij}\left[
\bigl\langle a_{i}^{\dagger }({p}_{2})a_{i}({p}_{1})\bigr\rangle _{{\bf e}}
\bigl\langle
a_{i}^{\dagger }({p}_{1})a_{i}({p}_{2})\bigr\rangle _{{\bf e}}-\bigl\langle
a_{i}^{\dagger }({p}_{1})\bigr\rangle _{{\bf e}}\bigl\langle a_{i}^{\dagger }({p}%
_{2})\bigr\rangle _{{\bf e}}\bigl\langle a_{i}^{{}}({p}_{1})\bigr\rangle _{{\bf e}}
\bigl\langle
a_{i}^{{}}({p}_{2})\bigr\rangle _{{\bf e}}\right].
\end{array}
\label{6}
\end{equation}
Here
\begin{equation}
\bigl\langle a_{i}^{\dagger }({p}_{1})a_{i}({p}_{2})\bigr\rangle _{{\bf e}}=
\bigl\langle
a_{i}^{\dagger }({p}_{1})a_{i}({p}_{2})\bigr\rangle _{ch}+
\bigl\langle a_{i}^{\dagger }({p}_{1})\bigr\rangle _{{\bf e}}
\bigl\langle a_{i}^{{}}({p}_{2})\bigr\rangle _{{\bf e}},
\label{7}
\end{equation}
where the irreducible (thermal) part of the two-operator average
\begin{equation}
\bigl\langle a_{i}^{\dagger }({p}_{1})a_{i}({p}_{2})\bigr\rangle _{ch}=\sqrt{%
p_{10}p_{20}}\bigl\langle {\rm a}_{i,qm}^{\dagger }({\bf p}_{1},t_{f}){\rm a}%
_{i,qm}({\bf p}_{2},t_{f})\bigr\rangle _{{\bf e}}  \label{8}
\end{equation}
does not depend on ${\bf e}$ \footnote{%
Such a dependence could take place if the mass shift were non-zero and
dependent on the ${\bf e}$-orientation of the quasi-pion vacuum.}
and
\begin{equation}
\bigl\langle a_{i}({p})\bigr\rangle _{{\bf e}}=
e_{i}d({p})\equiv e_{i}\sqrt{p_{0}%
}{\rm d}_{coh}({\bf p,}t_{f},t_{out})\text{ .}  \label{9}
\end{equation}

One can introduce the one-particle Wigner function \cite{Groot}
\begin{equation}
f_{{\bf e},i}(x,p)=(2\pi )^{-3}\int d^{4}q'\delta (q'\cdot
p)e^{iq'x}\bigl\langle a_{i}^{\dagger
}(p+q'/2)a_{i}(p-q'/2)\bigr\rangle _{{\bf e}},  \label{Wigner-f}
\end{equation}
satisfying the relation
\begin{equation}
p_{\mu }\partial ^{\mu }f_{{\bf e},i}(x,p)=0  \label{Wigner-f-tr}
\end{equation}
and describing the phase-space density
of the non-interacting pions at $t\geqslant $ $t_{out}$ or, in covariant
formalism, at $t\geqslant \sigma _{out}=t_{out}({\bf x})$; here $\sigma
_{out}$ is a space-time hypersurface where the
interactions are ''switched off''
and particles can be considered as free. From Eq. (\ref{Wigner-f}), we get
\begin{equation}
\bigl\langle a_{i}^{\dagger }({p}_{1})a_{i}({p}_{2})\bigr\rangle _{{\bf e}%
}=\int\limits_{\sigma _{out}}d\sigma _{\mu }p^{\mu }f_{{\bf e}%
,i}(x,p)e^{-iqx},~~q=p_{1}-p_{2},~~p=(p_{1}+p_{2})/2. \label{aver1}
\end{equation}

Using Eqs. (\ref{7}), (\ref{9}) and (\ref{aver1}), one can split
the Wigner function into the chaotic ($ch$) and coherent ($coh$)
components:
\begin{equation}
f_{{\bf e},i}(x,p)=f_{ch}(x,p)+|e_{i}|^{2}f_{coh}(x,p).  \label{11}
\end{equation}
Integrated over $\sigma _{out}$, these components determine
the operator averages
$\bigl\langle a_{i}^{\dagger }({p}_{1})a_{i}({p}_{2})\bigr\rangle_{ch}$ and
$\bigl\langle a_{i}^{\dagger }({p}_{1})\bigr\rangle _{{\bf e}}
\bigl\langle a_{i}^{{}}({p}_{2})\bigr\rangle _{{\bf e}}$ respectively:
\begin{equation}
\begin{array}{l}
\bigl\langle a_{i}^{\dagger }({p}_{1})a_{i}({p}_{2})\bigr\rangle
_{ch}=\int\limits_{\sigma _{out}}d\sigma _{\mu }p^{\mu }e^{-iq\cdot
x}f_{ch}(x,p),\\
\\
\bigl\langle a_{i}^{\dagger }({p}_{1})\bigr\rangle _{{\bf e}}
\bigl\langle a_{i}^{{}}({p}_{2})\bigr\rangle _{{\bf e}}=
|e_{i}|^{2}d^{\ast }({p}_{1})d({p}%
_{2})=|e_{i}|^{2}\int\limits_{\sigma _{out}}d\sigma _{\mu }p^{\mu
}e^{-iq\cdot x}f_{coh}(x,p).
\end{array}
\label{9a}
\end{equation}

We suppose that the system has zero{\it \ average} charge and
calculate the observables
averaging over the random orientation of the quasi-pion vacuum in the
isospin space ($d\Omega ({\bf e})=d\cos\theta d\phi$):
\begin{equation}
\bigl\langle ...\bigr\rangle \equiv Sp(...{\bf \rho )}=(4\pi )^{-1}{\bf %
\int }d\Omega ({\bf e})\langle \dots \rangle _{{\bf e}}\equiv (4\pi )^{-1}%
{\bf \int }d\Omega ({\bf e})Sp(\dots {\bf \rho }_{{\bf e}}).  \label{12}
\end{equation}
The {\it %
observable} pion field is related to the ensemble of events only, so
the corresponding complete averages of the asymptotically free operators
vanish, for example, $\bigl\langle a_{\pi ^{+}}({p})\bigr\rangle =
(4\pi )^{-1}{\bf \int }d\Omega ({\bf e)}\bigl\langle a_{\pi ^{+}}({p})
\bigr\rangle _{{\bf e}}=0$.
The averages of these operators also vanish for charge-constrained coherent
pion states $|c\rangle $, the states of a
fixed electric charge and isospin - so called generalized coherent states
\cite{Botke,GKW,Bhaumik}. This means that the density
matrix $\rho $ can be represented as a weighted sum of the
projection operators $|c\rangle \langle c|$
of these states.

To illustrate this statement, let us
consider a simple artificial case of only two sorts of
oppositely charged bosons in one mode.
Then the usual coherent states $\bigl| \alpha _{\lambda
}\bigr\rangle $, $\lambda =\pm $, are
\begin{equation}
\begin{array}{l}
\bigl| \alpha _{\lambda }\bigr\rangle =\exp (-\frac{1}{2}\left| \alpha
_{\lambda }\right| ^{2})\stackrel{\infty }{%
\mathrel{\mathop{\sum }\limits_{n=0}}%
}\frac{\alpha _{\lambda }^{n}}{(n!)^{1/2}}\bigl| n_{\lambda }\bigr\rangle
,\quad a_{\lambda }\bigl| \alpha _{\lambda }\bigr\rangle =\alpha _{\lambda
}\bigl| \alpha _{\lambda }\bigr\rangle , \\
\\
\bigl| n_{\lambda }\bigr\rangle =(n!)^{-1/2}(a_{\lambda }^{\dagger
})^{n}\bigl| 0_{\lambda }\bigr\rangle ,\quad \left[ a_{\lambda },a_{\lambda
^{\prime }}^{\dagger }\right] =\delta _{\lambda \lambda ^{\prime }},\quad
\alpha _{\pm }=\left| \alpha \right| e^{\pm i\phi }.
\end{array}
\label{13}
\end{equation}
These states represent superpositions of the states with
different charges and so violate the super-selection rule.
The charge-constrained
coherent state $|c_{0}\rangle $ of charged quanta
with a zero total charge may be
obtained by projecting this state out
from the charge-unconstrained two-component coherent state
$\bigl| \alpha _{+}\bigr\rangle
\bigl| \alpha _{-}\bigr\rangle $ \cite{Bhaumik}:
\begin{equation}
|c_{0}\rangle =\frac{1}{2\pi }\stackrel{2\pi }{%
\mathrel{\mathop{\int }\limits_{0}}%
}d\phi \bigl| \alpha _{+}\bigr\rangle \bigl| \alpha _{-}\bigr\rangle =\exp
(-\left| \alpha \right| ^{2})\stackrel{\infty }{%
\mathrel{\mathop{\sum }\limits_{n=0}}%
}\frac{\left| \alpha \right| ^{2n}}{n!}\bigl| n_{+}\bigr\rangle
\bigl|n_{-}\bigr\rangle .  \label{15}
\end{equation}
One may see that the zero charge state $|c_{0}\rangle $ represents a
superposition of the states with the same charges
(with equal numbers of particles and antiparticles)
and thus satisfies the super-selection rule.
Similarly, the density matrix
\begin{equation}
\begin{array}{c}
\widehat{\rho }=\frac{1}{2\pi }\stackrel{2\pi }{%
\mathrel{\mathop{\int }\limits_{0}}%
}d\phi \left| \alpha _{+}\bigr\rangle \left| \alpha _{-}\bigr\rangle
\bigl\langle \alpha _{+}\right| \bigl\langle \alpha _{-}\right| = \\
\\
\exp (-2\left| \alpha \right| ^{2})\stackrel{\infty }{%
\mathrel{\mathop{\sum }\limits_{n_{1}=0}}%
}\stackrel{\infty }{%
\mathrel{\mathop{\sum }\limits_{n_{2}=0}}%
}\stackrel{\infty }{%
\mathrel{\mathop{\sum }\limits_{n_{3}=0}}%
}\stackrel{\infty }{%
\mathrel{\mathop{\sum }\limits_{n_{4}=0}}%
}\frac{\left| \alpha \right| ^{n_{1}+n_{2}+n_{3}+n_{4}}}{%
(n_{1}!)^{1/2}(n_{2}!)^{1/2}(n_{3}!)^{1/2}(n_{4}!)^{1/2}}\delta
_{n_{1}-n_{2},n_{3}-n_{4}}\left| n_{1,+}\bigr\rangle \left|
n_{2,-}\bigr\rangle \bigl\langle n_{3,+}\right| \bigl\langle n_{4,-}\right|
\end{array}
\label{16}
\end{equation}
describes the mixture of the charge-constrained
coherent states $|c_{n}\rangle$:
\begin{equation}
\widehat{\rho }=\stackrel{\infty }{%
\mathrel{\mathop{\sum }\limits_{n=-\infty }}%
}|c_{n}\rangle \langle c_{n}|,  \label{18}
\end{equation}
where $|c_{n}\rangle$ is the coherent state of charge ''$n$'':
\begin{equation}
|c_{n}\rangle =\exp (-\left| \alpha \right| ^{2})\stackrel{\infty }{%
\mathrel{\mathop{\sum }\limits_{n_{1}=0}}%
}\stackrel{\infty }{%
\mathrel{\mathop{\sum }\limits_{n_{2}=0}}%
}\delta _{n_{1}-n_{2},n}\frac{\left| \alpha \right| ^{n_{1}+n_{2}}e^{i\phi
(n_{1}-n_{2})}}{(n_{1}!)^{1/2}(n_{2}!)^{1/2}}\bigl| n_{1,+}\bigr\rangle
\bigl| n_{2,-}\bigr\rangle .  \label{19}
\end{equation}
While, in our example, the system described by the density matrix $%
\widehat{\rho }$ has not a definite charge, the average charge is equal
to zero:
\begin{equation}
Sp(\widehat{\rho }(a_{+}^{\dagger }a_{+}-a_{-}^{\dagger }a_{-}))=0.
\label{20}
\end{equation}
Note, that the expectation values of the annihilation operators
in the corresponding coherent states are non-zero,
$\bigl\langle \alpha _{\lambda}\bigr|
a_{\lambda}\bigl| \alpha _{\lambda}\bigr\rangle =\alpha_{\lambda}$,
while $Sp(\widehat{\rho }a_{\lambda})=0$.

Continuing the discussion of coherent pion production, we will assume
the density matrix ${\bf \rho }_{{\bf e}}$ of a Gaussian-type
in terms of the quasi-particle annihilation (creation) operators
${\rm a}_{i,qm}({\bf p}{,}t_{f})$,
related to the free particle operators according to Eqs. (\ref{2})
and (\ref{3}). Then, similar to the above example, this density
matrix can be expressed through the projection operators on the usual
charge-unconstrained coherent states of free pion field.
Averaging ${\bf \rho }_{{\bf e}}$ over all directions of the
isovector ${\bf e}$ according to Eq.~(\ref{12}),
we finally get the density matrix $\rho $ in a form of a weighted
sum of the projection operators on the charge-constrained coherent
states describing, in agreement with the super-selection rule,
the system of a fixed average charge.\footnote{
We do not consider here the squeeze-states of the
density matrix conditioned by possible mass shift of quasi-particles.
Note, however, that
charged pions have anyway no squeeze-state components
\cite{Weiner}.
}

The expressions for pion spectra in Eq. (\ref{1}) thus contain
the averaging over the direction of the isovector ${\bf e}$.
As a result, the single-pion spectra are independent of pion
charges $i={\pm},{0}$:
\begin{equation}
\begin{array}{l}
\omega _{{\bf p}}\frac{d^{3}N_{i}}{d^{3}{\bf p}}=(4\pi )^{-1}{\bf \int }%
d\Omega ({\bf e)}\int d\sigma _{\mu }p^{\mu }f_{{\bf e},i}(x,p)=\int d\sigma
_{\mu }p^{\mu }f(x,p), \\
\\
f(x,p)=f_{ch}(x,p)+\frac{1}{3}f_{coh}(x,p),
\end{array}
\label{10}
\end{equation}
where we have used the equality $(4\pi )^{-1}{\bf \int }
d\Omega ({\bf e)}|e_{i}|^{2}=1/3.$
Note that the coherent part of the single--pion spectrum is
\begin{equation}
\omega _{{\bf p}}\frac{d^{3}N_{coh}}{d^{3}{\bf p}}\equiv
\omega _{{\bf p}}\frac{d^{3}N}{d^{3}{\bf p}}G(p)\equiv
\omega _{{\bf p}}\frac{d^{3}N_{ch}}{d^{3}{\bf p}}D(p)
=\frac{1}{3}\int
d\sigma _{\mu }p^{\mu }f_{coh}(x,p)=\frac{1}{3}|d({p})|^{2},
\label{21}
\end{equation}
where the functions $G(p)$ and $D(p)$ measure the coherent
fraction:
\begin{equation}
G(p)=\frac{D(p)}{1+D(p)}\equiv \frac{d^{3}N_{coh}/d^{3}{\bf p}}
{d^{3}N/d^{3}{\bf p}}
=\frac{\frac{1}{3}\int d\sigma _{\mu }p^{\mu }
f_{coh}(x,p)}{\int d\sigma _{\mu }p^{\mu}f(x,p)},~~
D(p)\equiv \frac{d^{3}N_{coh}/d^{3}{\bf p}}{d^{3}N_{ch}/d^{3}{\bf p}}
=\frac{\frac{1}{3}\int d\sigma _{\mu }p^{\mu }
f_{coh}(x,p)}{\int d\sigma _{\mu }p^{\mu}f_{ch}(x,p)}.  \label{23}
\end{equation}

The coherence influences also the quantum statistical (without FSI)
correlation functions:
\begin{equation}
C_{QS}^{ij}(p,q)=\frac{(4\pi )^{-1}{\bf \int }d\Omega ({\bf e)}\bigl\langle
a_{i}^{\dagger }({p}_{1})a_{j}^{\dagger }({p}_{2})a_{i}({p}_{1})a_{j}({p}%
_{2})\bigr\rangle _{{\bf e}}}
{\left( (4\pi )^{-1}{\bf \int }d\Omega ({\bf e)}%
\bigl\langle a_{i}^{\dagger }({p}_{1})a_{i}^{{}}({p}_{1})\bigr\rangle _{{\bf e}%
}\right) \left( (4\pi )^{-1}{\bf \int }d\Omega ({\bf e)}\bigl\langle
a_{j}^{\dagger }({p}_{2})a_{j}^{{}}({p}_{2})\bigr\rangle _{{\bf e}}\right)}.
\label{24}
\end{equation}
Taking into account Eqs. (\ref{6})-(\ref{9}), (\ref{23}) and the
equalities $p_{1,2}=p\pm q/2$, we get
\begin{equation}
\begin{array}{c}
C_{QS}^{ij}(p,q)=1+\bigl(9\bigl\langle |e_ie_j|^2\bigr\rangle - 1-
\delta_{ij}\bigr)G(p_1)G(p_2)+
\delta_{ij}\bigl\langle \cos (qx_{12})\bigr\rangle '\\
\\
=\bigl[1+D(p_1)\bigr]^{-1}\bigl[1+D(p_2)\bigr]^{-1}
\Bigl\{
1+D(p_1)+D(p_2)+9\bigl\langle |e_ie_j|^2\bigr\rangle
D(p_1)D(p_2)\\
\\
+\delta_{ij}\bigl\langle \cos (qx_{12})\bigr\rangle'_{ch}
\bigl[1+{\cal D}(p_1,p_2)+{\cal D}(p_2,p_1)\bigr]
\Bigr\},
\end{array}
\label{25}
\end{equation}
where the quasi--average
$\langle\cos(qx_{12})\rangle'\equiv\langle\cos(q(x_1-x_2))\rangle'$ is defined as:
\begin{equation}
\langle \cos (qx_{12})\rangle' =\frac{\int d^{3}\sigma _{\mu }(x_{1})d^{3}\sigma
_{\nu }(x_{2})p^{\mu }p^{\nu }f(x_{1},p)f(x_{2},p)\cos (qx_{12})}{%
\int d^{3}\sigma _{\mu }(x_{1})d^{3}\sigma _{\nu }(x_{2})p_{1}^{\mu
}p_{2}^{\nu }f(x_{1},p_{1})f(x_{2},p_{2})}  \label{26}
\end{equation}
and similarly, with the substitution $f\rightarrow f_{ch}$, the quasi--average
$\langle\cos(qx_{12})\rangle'_{ch}$;
the function
\begin{equation}
{\cal D}(p_1,p_2)=
\frac{\frac{1}{3}\int d\sigma _{\mu }p^{\mu }
f_{coh}(x,p)e^{-iq\cdot x}}
{\int d\sigma _{\mu }p^{\mu}f_{ch}(x,p)e^{-iq\cdot x}}=
\frac{\frac{1}{3}d^*(p_1)d(p_2)}
{\bigl\langle
a_{i}^{\dagger }({p}_{1})a_{i}({p}_{2})\bigr\rangle_{ch} },~~
{\cal D}(p,p)=D(p).
\label{d12}
\end{equation}
Note that
\begin{eqnarray}
\langle \cos (qx_{12})\rangle' &=&
G(p_1)G(p_2)+
\frac{1+{\cal D}(p_1,p_2)+{\cal D}(p_2,p_1)}
{[1+D(p_1)][1+D(p_2)]}
\langle \cos (qx_{12})\rangle'_{ch}
\nonumber \\
&& \label{26'} \\
&=&
\frac{1+{\cal D}(p_1,p_2)+{\cal D}(p_2,p_1)+
{\cal D}(p_1,p_2){\cal D}(p_2,p_1)}
{[1+D(p_1)][1+D(p_2)]}
\langle \cos (qx_{12})\rangle'_{ch}. \nonumber
\end{eqnarray}
Calculating the averages
\begin{equation}
\bigl\langle |e_{i}e_{j}|^{2}\bigr\rangle =
(4\pi )^{-1}{\bf \int }d\Omega ({\bf e)}%
|e_{i}e_{j}|^{2},
\label{27}
\end{equation}
\begin{equation}
\bigl\langle |e_{0}|^{4}\bigr\rangle =\frac{1}{5},~~
\bigl\langle |e_{\pm }|^{4}\bigr\rangle
=\bigl\langle |e_{+}e_{-}|^{2}\bigr\rangle =\frac{2}{15},~~
\bigl\langle |e_{0}e_{\pm
}|^{2}\bigr\rangle =\frac{1}{15},  \label{28}
\end{equation}
we get for the intercepts of the QS correlation functions:
\begin{equation}
\begin{array}{l}
C_{QS}^{++}(p,0)=2-\frac{4}{5}G^{2}(p),~~C_{QS}^{00}(p,0)=2-\frac{1}{5}%
G^{2}(p), \\
\\
C_{QS}^{+-}(p,0)=1+\frac{1}{5}G^{2}(p),~~C_{QS}^{+0}(p,0)=1-\frac{2}{5}%
G^{2}(p).
\end{array}
\label{inter}
\end{equation}
Particularly, it follows from Eqs. (\ref{inter}) that the decay of the
quasi-pion vacuum suppresses the correlation functions of identical charged
pions and enhances the one of non-identical charged pions, the latter effect
being by a factor of 4 smaller. For $G^{2}(p)=1$, the intercepts in Eqs.~(\ref
{inter}) coincide with those found in Ref. \cite{ggm93} in the case of a
strong pion condensation.
Our results however differ from the intercepts found
in the model \cite{Andreev,Nakamura} of pion
emission in a pure quantum state, - the charge-constrained
coherent state. They are
recovered only for large average numbers of coherent pions.
One can then replace the canonical ensemble corresponding to the pure quantum
state with a fixed charge, by the
grand canonical one, described by the density matrix of the ensemble with a fixed
{\it average} charge. For ultra-relativistic A+A collisions,
the inclusive description
based on the grand canonical ensemble is a fairly adequate approach,
allowing to built explicitly the density matrix for a mixture of thermal and
charge-constrained coherent radiations and make some calculations analytically.

One can check that the intercepts, as well as the QS correlation functions
at any $q$, satisfy the relation \cite{lyub}
\begin{equation}
C_{QS}^{++}+C_{QS}^{+-}=C_{QS}^{00}+C_{QS}^{+0}.  \label{QS-rel}
\end{equation}
This relation follows from the assumed isotopically unpolarized pion
emission. It is valid also for the complete correlation functions (with
FSI), except for the region of very small $|{\bf q}|$ where the correlation
functions of charged pions are strongly affected by the isospin
non-conserving Coulomb interaction.

Note that the correlation functions, as well as their QS parts, satisfy the
usual normalization condition $C(p,q)\rightarrow 1$ at large $|{\bf q}|$
provided that the coherent part of the Wigner density vanishes with the
increasing $|p\pm q/2|$ faster than the chaotic one, i.e. $G(p\pm
q/2)\rightarrow 0$ at large $|{\bf q}|$.

To get some insight in a possible behavior of the relative coherent
contribution $G(p)$, consider the situation when the system decays during
rather short time, $t_{out}-t_f\rightarrow 0$, and the partial (at a fixed $%
{\bf e}$) average of the pion annihilation operator has a simple Gaussian
form:
\begin{equation}
\bigl\langle a_{i}(p)\bigr\rangle _{{\bf e}}\sim \exp (-R_{coh}^{2}{\bf p}^{2}).
\label{25a}
\end{equation}
According to Eq. (\ref{9a}), the corresponding Wigner density
\begin{equation}
f_{coh}(x,p)\sim \exp (-2R_{coh}^{2}{\bf p}^{2}-{\bf x}^{2}/2R_{coh}^{2}),
\label{25b}
\end{equation}
so the parameter $R_{coh}$ determines not only the spectrum, but also the
characteristic radius of the region of the instantaneous coherent pion
emission in accordance with the minimized uncertainty relation $\Delta
x\Delta p=\hbar /2$. Let us assume a similar Gaussian parametrization of the
chaotic component of the Wigner density in the non-relativistic momentum
region:
\begin{equation}
f_{ch}(x,p)\sim \exp (-2R_{T}^{2}{\bf p}^{2}-{\bf x}^{2}/2R_{ch}^{2}),
\label{25c}
\end{equation}
where $R_{T}\equiv (4mT)^{-1/2}$ measures the characteristic size of the
single--pion emitter (heat de Broglie length) and $R_{ch}\geq R_{T}$ is the
characteristic radius of the region of the chaotic pion emission. In the
considered rare gas limit, we then get the correlator
\begin{equation}
\langle \cos (qx_{12})\rangle'_{ch}=\exp (-R^{2}{\bf q}^{2}),  \label{25d}
\end{equation}
where $R=(R_{ch}^{2}-R_{T}^{2})^{1/2}\approx R_{ch}$ represents (in the
absence of the coherent contribution) the usual interferometry radius. The
coherent fraction $G(p)=D(p)/[1+D(p)]$ and
\begin{equation}
D(p)=\frac{d^{3}N_{coh}/d^{3}{\bf p}}{d^{3}N_{ch}/d^{3}{\bf p}}\equiv \frac{%
\frac{1}{3}\int d\sigma _{\mu }p^{\mu }f_{coh}(x,p)}{\int d\sigma _{\mu
}p^{\mu }f_{ch}(x,p)}\sim \exp \left[ -2\left( R_{coh}^{2}-R_{T}^{2})\right)
{\bf p}^{2}\right] .
\label{rat-def}
\end{equation}
We see that $G(p)\rightarrow 0$ at large $|{\bf p}|$ on a reasonable
condition $R_{coh}>R_{T}$.

In fact, since the quasi--classical (coherent) pion emission is
conditioned by the decay of a thermal system, one may expect the effective radius
for the coherent radiation, $R_{coh}$, close to that for the thermal
emission, $R_{ch}$.
Generally, in dynamical models, the effective radius varies with
the momentum ${\bf p}$ and characterizes the size of the homogeneity region
- the region of a substantial density of the pions emitted at the
freeze--out time with three--momenta in the vicinity of ${\bf p}$. In this case,
both the coherent and chaotic radii practically coincide with the
homogeneity
length of the system. Assuming $R_{coh}\approx R_{ch}$, we have
${\cal D}(p_1,p_2)\approx {\cal D}(p,p)=D(p)$
and, according to Eq.~(\ref{26'}),
\begin{equation}
~~~\langle \cos (qx_{12})\rangle' \approx \frac{\lbrack 1+D(p)]^{2}}{%
[1+D(p+q/2)][1+D(p-q/2)]}\langle \cos (qx_{12})\rangle_{ch}'.
\label{25e}
\end{equation}
One can see that $\langle \cos (qx_{12})\rangle'
\approx \langle \cos(qx_{12})\rangle'
_{ch}$ at small $|{\bf q}|$ or, in the case of a small coherent contribution $%
D(p)\ll 1$. Note that in the opposite case, $D(p)\gg 1$,
a decrease of the correlation function towards unity with the increasing $%
{\bf q}^{2}$ is conditioned by the chaotic component $\langle \cos
(qx_{12})\rangle'_{ch}$ starting at ${\bf q}^{2}\sim R^{-2}\ln D^{2}({\bf 0})-4%
{\bf p}^{2}$. At smaller ${\bf q}^{2}$-values, the behavior of the
correlation function is essentially flatter due to the $q$-dependence of the
denominator in Eq. (\ref{25e}). For the extreme case of a pure coherent
radiation, $D(p)\rightarrow \infty $ ($G(p)\rightarrow 1$), the function
$\langle \cos (qx_{12})\rangle'$ tends to unity at all $q$
irrespective of the assumption $R_{coh}\approx R_{ch}$:
\begin{equation}
\langle \cos (qx_{12})\rangle' \rightarrow \frac{\int d^{3}\sigma _{\mu
}(x_{1})d^{3}\sigma _{\nu }(x_{2})p^{\mu }p^{\nu
}f_{coh}(x_{1},p)f_{coh}(x_{2},p)\cos (qx_{12})}{\int d^{3}\sigma _{\mu
}(x_{1})d^{3}\sigma _{\nu }(x_{2})p_{1}^{\mu }p_{2}^{\nu
}f_{coh}(x_{1},p_{1})f_{coh}(x_{2},p_{2})} = 1. \label{a1}
\end{equation}
The last equality in Eq. (\ref{a1}) follows from the definition
(\ref{9a}) of the coherent Wigner function,
both the nominator and denominator in Eq. (\ref{a1}) being equal to
$|d(p_1)d(p_2)|^2$.
Experimentally, the approach to such an extreme regime can display itself
as a tendency of the intercepts of the QS correlation functions
to the values defined by Eqs. (\ref{inter}) at $G(p)\rightarrow 1,$
and - as a flattering of the QS correlation functions within
a growing ${\bf q}$-interval. The latter mimics a decrease of
interferometry radii; of course, it does not mean that the real
size of the system tends to zero.

The effect of coherent radiation on pion spectra and $\pi ^{+}\pi ^{+}$ and $%
\pi ^{+}\pi ^{-}$ correlation functions is demonstrated in Figs. 1-3 for
different ratios $D_{tot}=D({\bf 0})(R_{T}/R_{coh})^{3}$ of the total numbers of
coherent and chaotic pions. The plots correspond to simple Gaussian Wigner
functions (\ref{25b}), (\ref{25c}) with $R_{T}\equiv (4mT)^{-1/2}\approx
0.72 $ fm ($T=0.135$ GeV) and $R_{coh}=R_{ch}=5$ fm. Under the assumption of
a common source of coherent and chaotic pions in ultra--relativistic heavy
ion collisions, characterized by a typical radius $R\sim 5-10$ fm, the
coherent component in the spectra is concentrated in rather small momentum
region of a characteristic width $(2R)^{-1}\sim 20-10$ MeV/c (see Fig. 1).

\begin{figure}
\centerline{\hskip -0.cm
\epsfig{width=15.cm,clip=1,figure=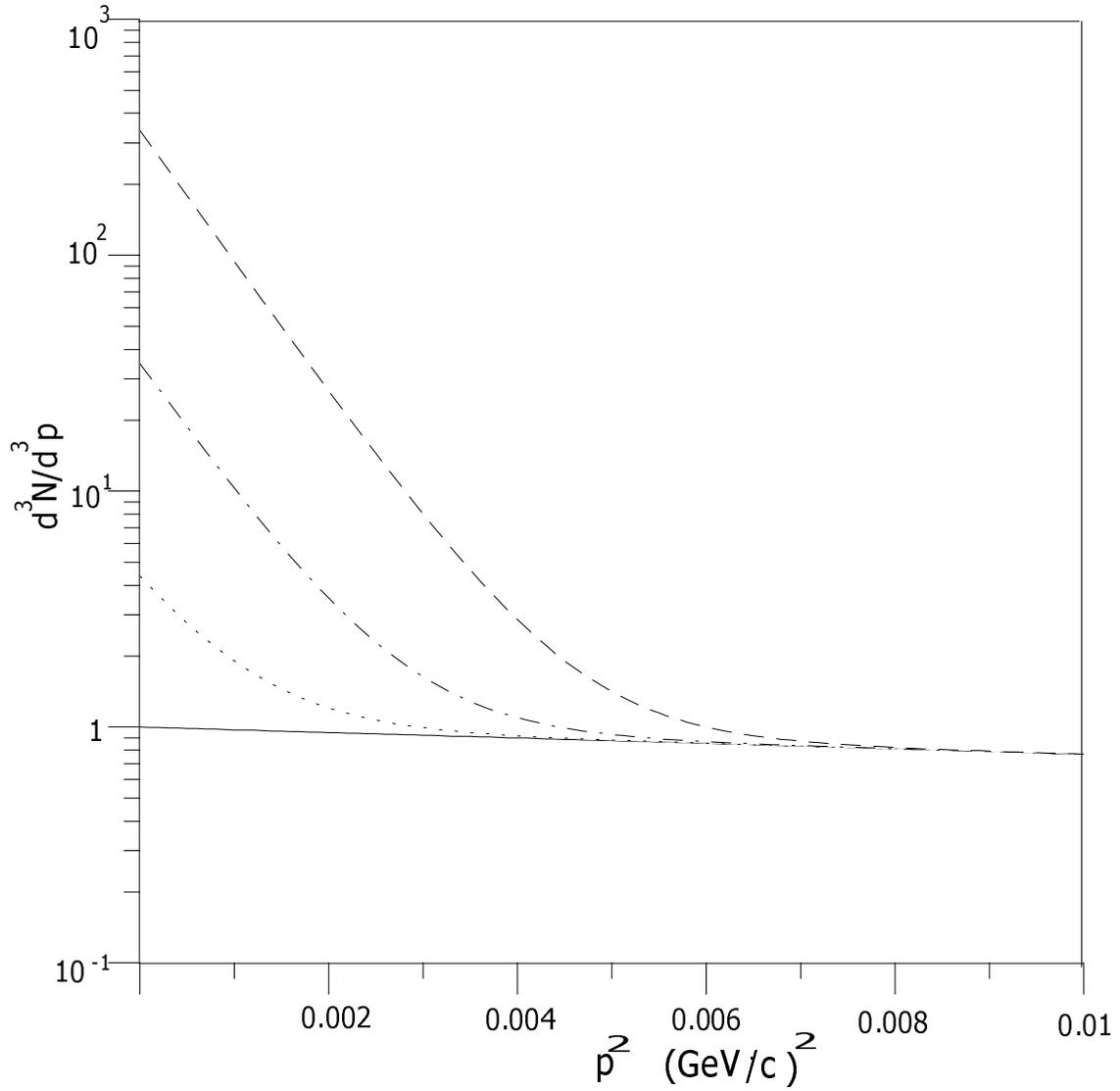} } \vspace*{-0.2cm}

\medskip
\caption{ The single-pion momentum spectra $d^{3}N/d^{3}{\bf p}$
calculated for different ratios $D_{tot}$ of the total numbers of
coherent and chaotic pions, assuming the Gaussian parametrization
of the Wigner densities in Eqs. (\ref{25b}), (\ref{25c}) with
$R_{T}\equiv (4mT)^{-1/2} \approx 0.72$ fm ($T=0.135$ GeV) and
$R_{coh}=R_{ch}=5$ fm. The solid, dotted, dash-dotted and dashed
curves correspond to $D_{tot}=0$, $0.01$, $0.1$, and $1$ respectively.
The overall normalization is arbitrary. }
\end{figure}

\begin{figure}
\centerline{\hskip -0.cm
\epsfig{width=15.cm,clip=1,figure=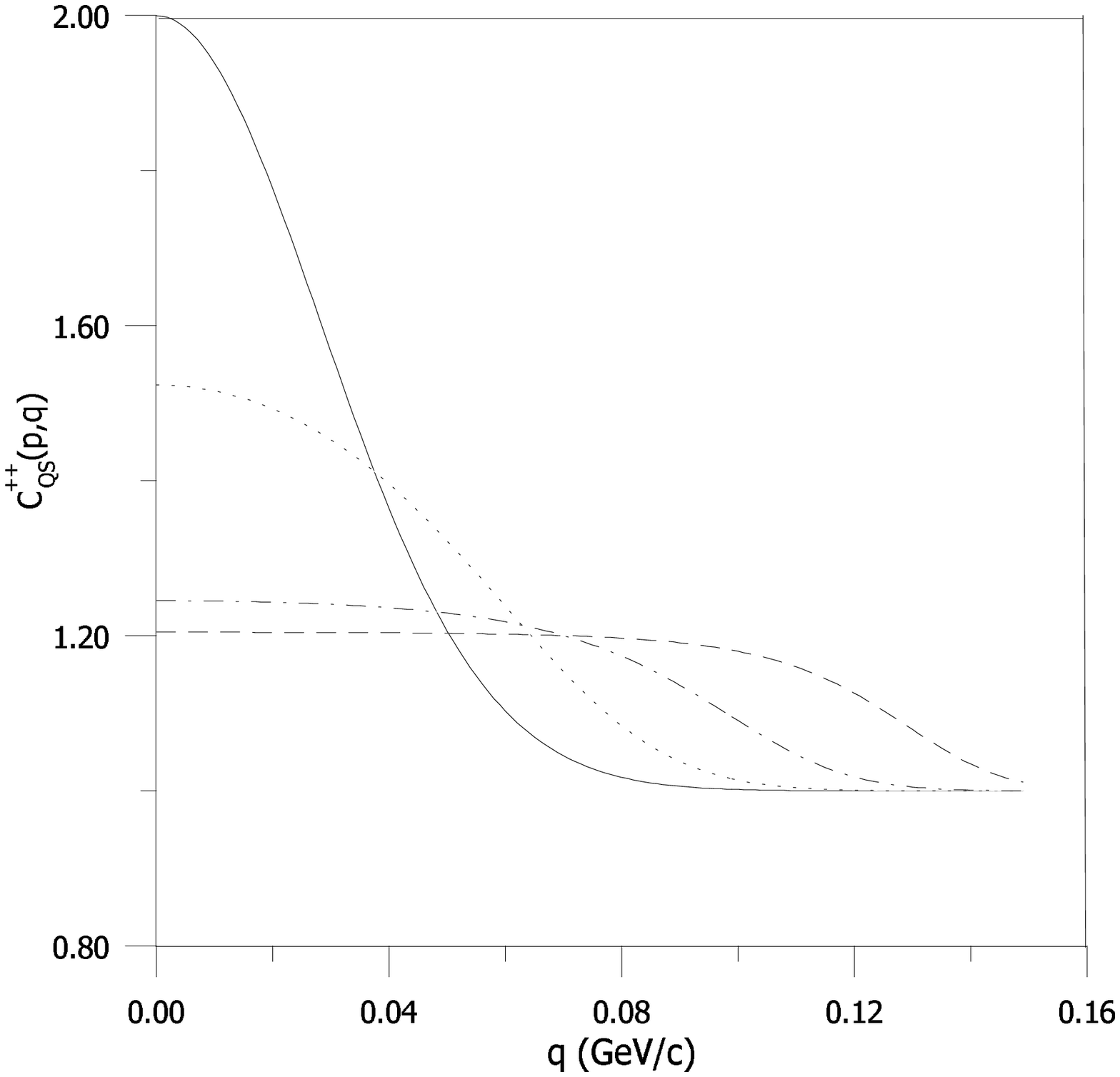} } \vspace*{-0.2cm}

\medskip
\caption{ The pure QS correlation functions $C_{QS}(p,q)$
calculated for $\pi ^{+}\pi ^{+}$ pairs at ${\bf p}={\bf 0}$ GeV/c
on the same conditions as in Fig. 1. }
\end{figure}

\begin{figure}
\centerline{\hskip -0.cm
\epsfig{width=15.cm,clip=1,figure=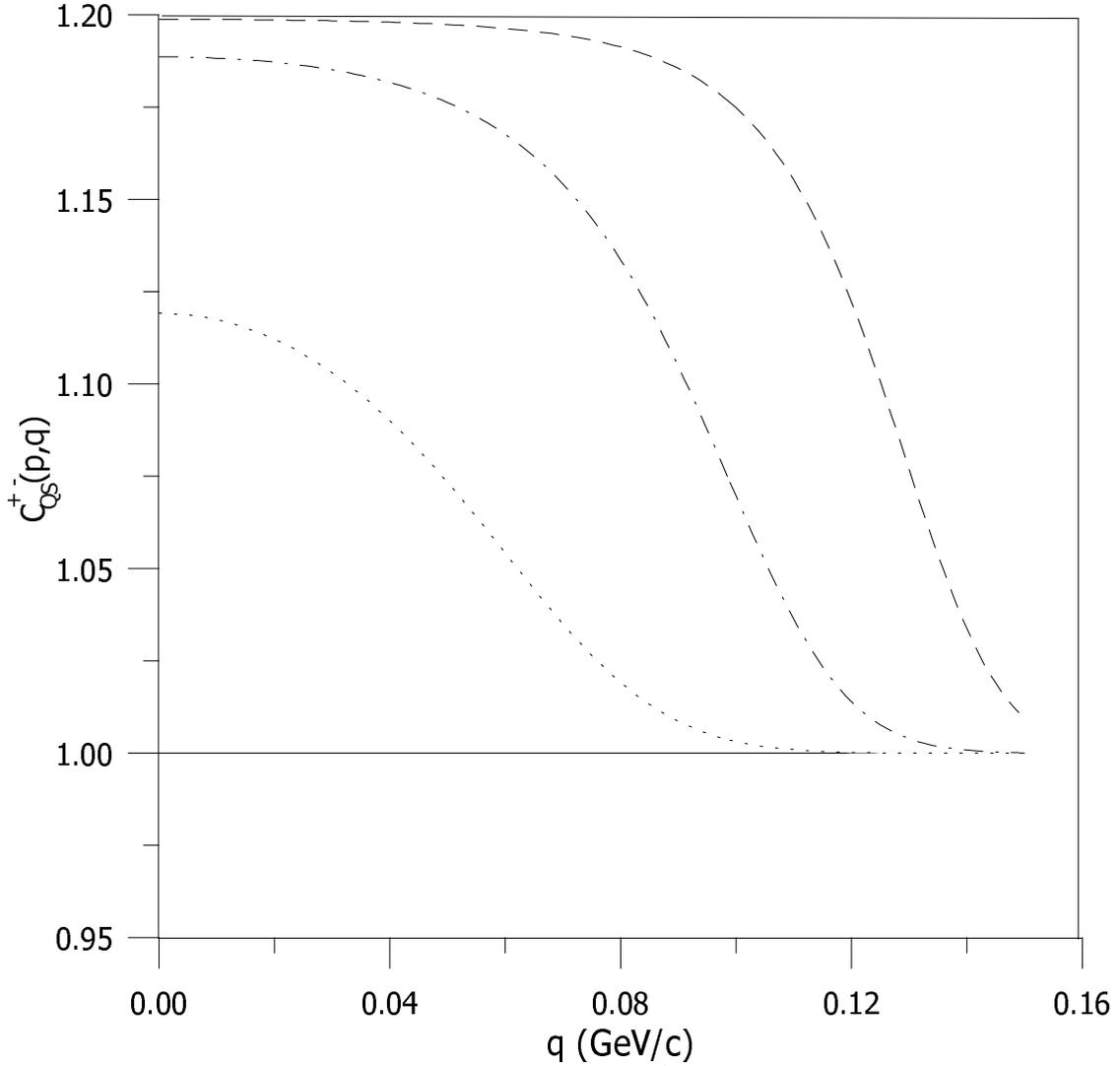} } \vspace*{-0.2cm}

\medskip
\caption{ The pure QS correlation functions $C_{QS}(p,q)$
calculated for $\pi ^{+}\pi ^{-}$ pairs at ${\bf p}={\bf 0}$ GeV/c
on the same conditions as in Fig. 1. }
\end{figure}

\section{Correlation functions affected by final state interaction and
coherence}

In ultrarelativistic A+A collisions, free hadrons appear mainly at the
late stage of the evolution after the system expands and reaches the thermal
freeze-out. After the hydrodynamic tube decays and produces final particles
and resonances, particles still appear from resonance decays.
Thus, more than half of pions produced in high energy heavy ion
collisions is of the resonance origin. As
a consequence, the pion spectra and correlations
are influenced by resonance production and decay
spectra, as well as - by resonance lifetimes. Particularly,
the pions from the decays of long-lived resonances do not contribute
to QS and FSI correlations and thus suppress the
correlation function $C^{ij}(p,q)$;
we will consider this suppression in next Section.

However, even after the thermal (hydrodynamic) system and short--lived
resonances decay, the particles in near--by phase space points continue
to interact.
Due to a large effective emission volume in
heavy ion collisions, the particle interaction in the final state
is usually dominated by the long-range Coulomb forces.
To calculate the
FSI effect on two-particle spectra, we will assume sufficiently small phase
space density of the produced particles and use the FSI theory in the
two--body approximation \cite{GKW,Led1,Lednicky} for pions, neglecting the
FSI of resonances.

The single--pion spectrum in Eq. (\ref{1}) then remains unchanged while the
two--pion one (for pairs containing no pions from long--lived sources) takes
the form
\begin{equation}
\omega _{{\bf p}_{1}}\omega _{{\bf p}_{2}}\frac{d^{6}N_{ij}}{d^{3}{\bf p}%
_{1}d^{3}{\bf p}_{2}}\doteq
\int d^4k_1d^4k_2d^4k_1'd^4k_2'\bigl\langle a_i^{\dagger }(k_1)
a_j^{\dagger }(k_2)a_i(k_1')a_j(k_2')\bigr\rangle
\Phi_{p_1p_2}^{(-)ij}(k_1,k_2)\Phi_{p_1p_2}^{(-)ij*}(k_1',k_2'),
\label{a4}
\end{equation}
where the non--symmetrized Bethe--Salpeter amplitude
$\Phi_{p_1p_2}^{(-)ij}(k_1,k_2)\equiv \Phi_{p_1p_2}^{(+)ij*}(k_1,k_2)$
in four--momentum representation is expressed through
the propagators of particles $i$ and $j$ and their scattering amplitude
${\cal F}_{ij}$ analytically continued to the unphysical region
\cite{Led1,Lednicky}:\footnote
{
It is important that the relation between the production amplitude and the
operator product average, as given in Eq. (\ref{1}), is valid also off mass
shell.
}
\begin{equation}
\Phi_{p_1p_2}^{(-)ij}(k_1,k_2)=\delta^4(k_1-p_1)\delta^4(k_2-p_2)+
\delta^4(k_1+k_2-p_1-p_2)
\frac{i\sqrt{p^{2}}}{\pi ^{3}}\frac{{\cal F}_{ij}^*
(k_1,k_2;p_1,p_2)}{(k_1^2-m^2-i0)(k_2^2-m^2-i0)}.
\label{a5}
\end{equation}
\noindent
The averaging in Eq.~(\ref{a4}) is performed with the help
of the statistical operator ${\bf \rho }$ without FSI:
$\langle \dots\rangle =Sp(\dots{\bf \rho })$.
Introducing the Bethe-Salpeter amplitudes
$\Psi_{p_1p_2}^{(-)ij}(x_1,x_2)$ in space-time representation:
\begin{equation}
\Phi_{p_1p_2}^{(-)ij}(k_1,k_2)=(2\pi)^{-8}\int d^4x_1d^4x_2
e^{ik_1x_1+ik_2x_2}\Psi_{p_1p_2}^{(-)ij}(x_1,x_2),
\label{bsft}
\end{equation}
one can rewrite Eq.~(\ref{a4}) as
\begin{equation}
\omega _{{\bf p}_{1}}\omega _{{\bf p}_{2}}\frac{d^{6}N_{ij}}{d^{3}{\bf p}%
_{1}d^{3}{\bf p}_{2}}\doteq
\int d^4x_1d^4x_2d^4x_1'd^4x_2'
\rho^{ij}(x_1,x_2;x_1',x_2')
\Psi_{p_1p_2}^{(-)ij}(x_1,x_2)\Psi_{p_1p_2}^{(-)ij*}(x_1',x_2'),
\label{a4'}
\end{equation}
where the space--time density matrix $\rho^{ij}$ is just the
Fourier transform of the four--operator average in
Eq.~(\ref{a4}):\footnote
{For identical particles, it differs from the space--time density matrix
of ref.~\cite{Led1}, where the effect of QS enters
through the symmetrization of the Bethe--Salpeter amplitudes
while, here - through the Wigner decomposition
of the four--operator average in Eq.~(\ref{a6}) below.
}
\begin{equation}
\rho^{ij}(x_1,x_2;x_1',x_2')=(2\pi)^{-16}
\int d^4k_1d^4k_2d^4k_1'd^4k_2'
e^{ik_1x_1+ik_2x_2}e^{-ik_1'x_1'-ik_2'x_2'}
\bigl\langle a_i^{\dagger }(k_1)
a_j^{\dagger }(k_2)a_i(k_1')a_j(k_2')\bigr\rangle .
\label{stdm}
\end{equation}
Separating the phase factor due to free motion of the
two--particle c.m.s.:
\begin{equation}
\begin{array}{c}
\Psi_{p_1p_2}^{(-)ij}(x_1,x_2)=
e^{-iPX_{12}}\psi_{q}^{(-)ij}(x_{12}),\\
\\
X_{12}=\frac12(x_1+x_2),~~x_{12}=x_1-x_2,~~P\equiv 2p=p_1+p_2
\label{bsr}
\end{array}
\end{equation}
and integrating over the pair c.m.s. four--coordinates
$X_{12}$ and $X_{12}'$ in Eq.~(\ref{a4'}), one can express
the two--particle spectrum through the reduced
space--time density matrix $\rho_P^{ij}(x_{12};x_{12}')$, the
latter depending on the pair total four--momentum $P$ and the
relative four--coordinates of the emission points only:
\begin{equation}
\omega _{{\bf p}_{1}}\omega _{{\bf p}_{2}}\frac{d^{6}N_{ij}}{d^{3}{\bf p}%
_{1}d^{3}{\bf p}_{2}}\doteq
\int d^4x_{12}d^4x_{12}'
\rho_P^{ij}(x_{12};x_{12}')
\psi_{q}^{(-)ij}(x_{12})\psi_{q}^{(-)ij*}(x_{12}'),
\label{a4''}
\end{equation}
\begin{equation}
\rho_P^{ij}(x_{12};x_{12}')=(2\pi)^{-8}
\int d^4k_1d^4k_1'
e^{i(k_1-p)x_{12}}e^{-i(k_1'-p)x_{12}'}
\bigl\langle a_i^{\dagger }(k_1)
a_j^{\dagger }(P-k_1)a_i(k_1')a_j(P-k_1')\bigr\rangle .
\label{stdm'}
\end{equation}
Note that in the two--particle c.m.s., where
$P=\{m_{12},0,0,0\}$, $q=\{0,2{\rm k}^*\}$,
$x_{12}=\{t^*,{\rm r}^*\}$, the reduced Bethe--Salpeter amplitude
$\psi_{q}^{(-)ij*}(x_{12})=\psi_{q}^{(+)ij}(x_{12})$
at $t^*=t_1^*-t_2^*=0$ coincides with a stationary
solution $\psi_{-{\rm k}^*}({\rm r}^*)$ of the scattering
problem having at large distances $r^*$ the asymptotic form
of a superposition of of the plane and outgoing spherical
waves (the minus sign of the vector ${\rm k}^*$ corresponds
to the reverse in time direction of the emission process).
This amplitude can be substituted by this solution
({\it equal time} approximation) on condition \cite{Lednicky}
$|t^*|\ll mr^{*2}$ which is usually satisfied for particle
production in heavy ion collisions.

Since the resonances have finite lifetimes, their decay products are created
in an essentially four--dimensional space-time region.
At the post thermal freeze-out stage, the
resonances are usually described by semiclassical techniques;
they are considered as unstable particles moving along classical trajectories
and decaying according to the exponential law \cite{Bolz}
(see, however, \cite{Led1,lp92,cp01}).
This approximation neglects a small correlation effect
in pairs of unlike pions appearing due to QS
correlations of identical resonances.
The resonances are supposed to be described according to the Gibbs
density matrix prior to the thermal freeze-out;
this guarantees the chaoticity of the decay pions.\footnote
{
Note that the chaotisation of decay pions partially happens
irrespective of the form of the density matrix if pions were emitted by
a large number of many different sorts of resonances.
}
Therefore, the pions from resonance decays do not
destroy the structure of the decomposition of the operator averages
in Eqs. (\ref{6}) and (\ref{7})
into irreducible parts based on the thermal Wick theorem.

After the production, the pions in near-by phase space points,
chaotic as well as coherent ones, undergo a long-time scale interaction
in the final state. According to Eqs.~(\ref{a4'}) or (\ref{a4''}),
the intensity of FSI interaction is conditioned by the
two--particle Bethe--Salpeter amplitudes $\Psi_{p_1p_2}(x_1,x_2)$ or
$\psi_q(x_{12})$ and the corresponding two--particle space--time density
matrices $\rho(x_1,x_2;x_1',x_2')$ or $\rho_P(x_{12};x_{12}')$.
Clearly, in the case of absent FSI, the two--pion spectrum
merely reduces to the Fourier transform of the space--time density
matrix. It can be represented as an integral over the mean
four--coordinates $\bar{x}=(x+x')/2$ of a combination of bilinear products of
single--particle chaotic and coherent emission functions $g_{ch}(\bar{x},p)$
and $g_{coh}(\bar{x},p)$, respectively defined in Eqs. (\ref{aa-g-1}) and
(\ref{aa-g-2}) below.

The emission function $g(\bar{x},p)$ is closely related with
the Wigner phase space density $f(x,p)$
at asymptotic times $t\geqslant t_{out}$.
Let us denote by $\bar{x}\equiv \{\bar{t},{\bf x}-({\bf p}/p_{0})(t-\bar{t})\}$
the space-time point, starting from which a free particle moving with
velocity $p/p_0$ reaches a point $x$;
the portion of such particles is $g(\bar{x},p)$.
Collecting all the contributions (starting in our case from the
thermal freeze-out time $t_{f}$), we have
\begin{equation}
p_{0}f(x,p)=\int d^4\bar{x}
\delta^3\bigl(\bar{{\bf x}}-{\bf x}+({\bf p}/p_{0})(t-\bar{t})\bigr)
g(\bar{x},p),
\label{f-out}
\end{equation}
where $g(\bar{x},p)=p_{0}\delta (\bar{t}-t_{f})f(\bar{x},p)+
s(\bar{x},p)$ and $s(\bar{x},p)\sim \theta (t-\bar{t})
\theta (\bar{t}-t_{f})$ is the density of pion emission at the
post--thermal stage, $t>t_{f}$. Therefore we can rewrite the irreducible
(thermal) part of the two-operator average through the chaotic emission
function as:
\begin{equation}
\begin{array}{c}
\bigl\langle a_{i}^{\dagger }({p}_{1})a_{i}({p}_{2})\bigr\rangle
_{ch}=
\int\limits_{\sigma _{out}}d\sigma _{\mu }p^{\mu
}e^{-iqx}f_{ch}(x,p)=
\int d^{4}\bar{x}e^{-iq\bar{x}}g_{ch}\left(\bar{x},p\right),\\
\\
p\equiv P/2=(p_1+p_2)/2,~~ q=p_1-p_2,
\end{array}
\label{aa-g-1}
\end{equation}
where we have used the equality
$qx=q\bar{x}$ following from the relation $qp\equiv q_0p_0-{\bf qp}=0$.
Similarly, for the coherent component of the two-operator average
at fixed ${\bf e}$, we get
\begin{equation}
\begin{array}{c}
\bigl\langle a_{i}^{\dagger }({p}_{1})\bigr\rangle _{{\bf e}}
\bigl\langle a_{i}^{{}}({p}_{2})\bigr\rangle _{{\bf e}}=
|e_{i}|^2d^*(p_1))d(p_2)=\\
\\
|e_{i}|^2\int\limits_{\sigma _{out}}d\sigma _{\mu }p^{\mu
}e^{-iqx}f_{coh}(x,p)=|e_{i}|^2\int d^{4}\bar{x}e^{-iq\bar{x}}
g_{coh}\left(\bar{x},p\right).
\end{array}
\label{aa-g-2}
\end{equation}
The results of Section II can thus be rewritten in terms
of the emission functions in accordance with a formal
substitution
$\int\limits_{\sigma _{out}}d\sigma _{\mu }p^{\mu}f(x,p)
\rightarrow \int d^{4}x g\left(x,p\right)$.

To express the four-operator average in Eq. (\ref{stdm'})
through the emission functions, we can exploit the decomposition
similar to that in Eq.~(\ref{6}):
\begin{equation}
\begin{array}{c}
\bigl\langle a_{i}^{\dagger }({k}_1)a_{j}^{\dagger }(P-k_1)
a_{i}(k_1')a_{j}(P-k_1')\bigr\rangle _{{\bf e}}=
\bigl\langle a_{i}^{\dagger }(k_1)a_{i}(k_1')\bigr\rangle _{{\bf e}}
\bigl\langle
a_{j}^{\dagger }(P-k_1)a_{j}(P-k_1')\bigr\rangle _{{\bf e}}+\\
\\ \delta_{ij}\left[
\bigl\langle a_{i}^{\dagger }(k_1)a_{i}(P-k_1')\bigr\rangle _{{\bf e}}
\bigl\langle a_{i}^{\dagger }(P-k_1)a_{i}(k_1')\bigr\rangle _{{\bf e}}
-\bigl\langle
a_{i}^{\dagger }(k_1)\bigr\rangle _{{\bf e}}\bigl\langle a_{i}^{\dagger
}(P-k_1)\bigr\rangle _{{\bf e}}\bigl\langle a_{i}(k_1')\bigr\rangle _{{\bf e}}
\bigl\langle a_{i}(P-k_1')\bigr\rangle _{{\bf e}}\right].
\end{array}
\label{a6}
\end{equation}
Using Eqs.~(\ref{aa-g-1}) and (\ref{aa-g-2}) for the two-operator averages
in Eq.~(\ref{a6}), we get:
\begin{equation}
\begin{array}{l}
\bigl\langle a_{i}^{\dagger }(k_1)a_{j}^{\dagger }(P-k_1)
a_{i}(k_1')a_{j}(P-k_1')\bigr\rangle _{{\bf e}}=
\int d^{4}\bar{x}_{1}d^{4}\bar{x}_{2}\times \\
\\
\Bigl\{e^{-i(k_{1}-k_{1}')\cdot \bar{x}_{12}}
g_{{\bf e},i}\left(\bar{x}_{1},\frac12(k_{1}+k_{1}')\right)
g_{{\bf e},j}\left(\bar{x}_{2},P-\frac12(k_{1}+k_{1}')\right)
+ \Bigr. \\
\\
\delta _{ij}e^{-i(k_{1}+k_{1}'-P)\cdot \bar{x}_{12}}
\bigl[g_{{\bf e},i}\left(\bar{x}_{1},p+\frac12(k_{1}-k_{1}')\right)
g_{{\bf e},i}\left(\bar{x}_{2},p-\frac12(k_{1}-k_{1}')\right) \bigr. \\
\\ \Bigl. \bigl.
-| e_{i}| ^{4}
g_{coh}\left(\bar{x}_{1},p+\frac12(k_{1}-k_{1}')\right)
g_{coh}\left(\bar{x}_{2},p-\frac12(k_{1}-k_{1}')\right)\bigr]\Bigr\},
\end{array}
\label{a27.1}
\end{equation}
where $\bar{x}_{12}=\bar{x}_1-\bar{x}_2$ and
\begin{equation}
g_{{\bf e},i}(\bar{x},k)=g_{ch}(\bar{x},k)+|e_{i}|^{2}g_{coh}(\bar{x},k).
\label{g}
\end{equation}
After the averaging over the orientation of the isospin
vector ${\bf e}$, we get
\begin{equation}
\begin{array}{l}
\bigl\langle a_{i}^{\dagger }({k}_{1})a_{j}^{\dagger }(P-k_1)
a_{i}({k}_{1}')a_{j}(P-k_1')\bigr\rangle =
\int d^{4}\bar{x}_{1}d^{4}\bar{x}_{2}
\cdot\\
\\
\Bigl\{e^{-i(k_{1}-k_{1}')\cdot \bar{x}_{12}}
\bigl[g\left(\bar{x}_{1},\frac12(k_{1}+k_{1}')\right)
g\left(\bar{x}_{2},P-\frac12(k_{1}+k_{1}')\right) \bigr.\Bigr. \\
\\ \bigr.
+\bigl(\bigl\langle | e_{i}e_{j}| ^{2}\bigr\rangle -\frac{1}{9}\bigr)
g_{coh}\left(\bar{x}_{1},\frac12(k_{1}+k_{1}')\right)
g_{coh}\left(\bar{x}_{2},P-\frac12(k_{1}+k_{1}')\right) \bigr]+ \\
\\
\delta _{ij}e^{-i(k_{1}+k_{1}'-P)\cdot \bar{x}_{12}}
\bigl[g\left(\bar{x}_{1},p+\frac12(k_{1}-k_{1}')\right)
g\left(\bar{x}_{2},p-\frac12(k_{1}-k_{1}')\right) \bigr. \\
\\ \Bigl.\bigl.
-\frac{1}{9}g_{coh}\left(\bar{x}_{1},p+\frac12(k_{1}-k_{1}')\right)
g_{coh}\left(\bar{x}_{2},p-\frac12(k_{1}-k_{1}')\right)\bigr]\Bigr\},
\end{array}
\label{b1}
\end{equation}
where
\begin{equation}
g(\bar{x},k)=g_{ch}(\bar{x},k)+\frac{1}{3}g_{coh}(\bar{x},k).
\label{b2}
\end{equation}
Inserting expression (\ref{b1}) for the four--operator average
into Eq.~(\ref{stdm'}) and,
integrating in the first and second term over $(k_1-k_1')$ and
$(k_1+k_1'-P)$ respectively, one can rewrite the reduced
space--time density matrix as:
\begin{equation}
\begin{array}{l}
\rho_P^{ij}(x_{12};x_{12}')=(2\pi)^{-4}
\int d^{4}\bar{x}_{1}d^{4}\bar{x}_{2}d^4\kappa
\cdot\\
\\
\Bigl\{e^{i\kappa\cdot(x_{12}-x_{12}')}
\delta^4\bigl(\frac12(x_{12}+x_{12}')-\bar{x}_{12}\bigr)
\bigl[g\left(\bar{x}_{1},p+\kappa\right)
g\left(\bar{x}_{2},p-\kappa\right) \bigr.\Bigr. \\
\\ \bigr.
+\bigl(\bigl\langle | e_{i}e_{j}| ^{2}\bigr\rangle -\frac{1}{9}\bigr)
g_{coh}\left(\bar{x}_{1},p+\kappa\right)
g_{coh}\left(\bar{x}_{2},p-\kappa\right) \bigr]+ \\
\\
\delta _{ij}e^{i\kappa\cdot(x_{12}+x_{12}')}
\delta^4\bigl(\frac12(x_{12}-x_{12}')-\bar{x}_{12}\bigr)
\bigl[g\left(\bar{x}_{1},p+\kappa\right)
g\left(\bar{x}_{2},p-\kappa\right) \bigr. \\
\\ \Bigl.\bigl.
-\frac{1}{9}g_{coh}\left(\bar{x}_{1},p+\kappa\right)
g_{coh}\left(\bar{x}_{2},p-\kappa\right)\bigr]\Bigr\}.
\end{array}
\label{stdm''}
\end{equation}
According to Eq.~(\ref{a4''}) and using
the equality $\psi_q(-\bar{x}_{12})=\psi_{-q}(\bar{x}_{12})$,
the two--pion spectrum then
becomes:
\begin{equation}
\begin{array}{c}
\omega _{{\bf p}_{1}}\omega _{{\bf p}_{2}}\frac{d^{6}N_{ij}}{d^{3}{\bf p}%
_{1}d^{3}{\bf p}_{2}}\doteq (2\pi)^{-4}
\int d^{4}\bar{x}_{1}d^{4}\bar{x}_{2}
d^4\kappa d^4\epsilon\, e^{i\kappa\cdot\epsilon}
\cdot\\
\\
\Bigl\{
\bigl[g\left(\bar{x}_{1},p+\kappa\right)
g\left(\bar{x}_{2},p-\kappa\right)
+\bigl(\bigl\langle | e_{i}e_{j}| ^{2}\bigr\rangle -\frac{1}{9}\bigr)
g_{coh}\left(\bar{x}_{1},p+\kappa\right)
g_{coh}\left(\bar{x}_{2},p-\kappa\right) \bigr]
\psi_{q}^{(-)ij}(\bar{x}_{12}+\frac12\epsilon)
\psi_{q}^{(-)ij*}(\bar{x}_{12}-\frac12\epsilon) \Bigr. \\
\\ \Bigl.\bigl.
+\delta _{ij}
\bigl[g\left(\bar{x}_{1},p+\kappa\right)
g\left(\bar{x}_{2},p-\kappa\right)
-\frac{1}{9}g_{coh}\left(\bar{x}_{1},p+\kappa\right)
g_{coh}\left(\bar{x}_{2},p-\kappa\right)\bigr]
\psi_{q}^{(-)ij}(\bar{x}_{12}+\frac12\epsilon)
\psi_{-q}^{(-)ij*}(\bar{x}_{12}-\frac12\epsilon)
\Bigr\}\\
\\
= (2\pi)^{-4}
\int d^{4}\bar{x}_{1}d^{4}\bar{x}_{2}
d^4\kappa d^4\epsilon\, e^{i\kappa\cdot\epsilon}
\cdot\\
\\
\Bigl\{
\bigl[
g_{ch}\left(\bar{x}_{1},p+\kappa\right)
g_{ch}\left(\bar{x}_{2},p-\kappa\right)+
\frac13\bigl(
g_{ch}\left(\bar{x}_{1},p+\kappa\right)
g_{coh}\left(\bar{x}_{2},p-\kappa\right)+
g_{coh}\left(\bar{x}_{1},p+\kappa\right)
g_{ch}\left(\bar{x}_{2},p-\kappa\right)\bigr)\bigr]\cdot\Bigr. \\
\\ \Bigl.
\bigl[
\psi_{q}^{(-)ij}(\bar{x}_{12}+\frac12\epsilon)
\psi_{q}^{(-)ij*}(\bar{x}_{12}-\frac12\epsilon)+
\delta_{ij}\psi_{q}^{(-)ij}(\bar{x}_{12}+\frac12\epsilon)
\psi_{-q}^{(-)ij*}(\bar{x}_{12}-\frac12\epsilon)\bigr]\Bigr. \\
\\ \Bigl.
+\langle | e_{i}e_{j}| ^{2}\bigr\rangle
g_{coh}\left(\bar{x}_{1},p+\kappa\right)
g_{coh}\left(\bar{x}_{2},p-\kappa\right)
\psi_{q}^{(-)ij}(\bar{x}_{12}+\frac12\epsilon)
\psi_{q}^{(-)ij*}(\bar{x}_{12}-\frac12\epsilon)
\Bigr\}.
\label{a4n}
\end{array}
\end{equation}
If the FSI were absent, i.e.
$\psi_{q}^{(-)ij}(\bar{x}_{12})=\exp(-iq\cdot\bar{x}_{12}/2)$, one would get
\begin{equation}
\begin{array}{c}
\omega _{{\bf p}_{1}}\omega _{{\bf p}_{2}}\frac{d^{6}N_{ij}}{d^{3}{\bf p}%
_{1}d^{3}{\bf p}_{2}}\doteq
\int d^{4}\bar{x}_{1}d^{4}\bar{x}_{2}\,
\Bigl\{
g\left(\bar{x}_{1},p_1\right)g\left(\bar{x}_{2},p_2\right)
+\bigl(\bigl\langle | e_{i}e_{j}| ^{2}\bigr\rangle -\frac{1}{9}\bigr)
g_{coh}\left(\bar{x}_{1},p_1\right)
g_{coh}\left(\bar{x}_{2},p_2\right) \Bigr. \\
\\ \Bigl.\bigl.
+\delta _{ij}
\bigl[g\left(\bar{x}_{1},p\right)g\left(\bar{x}_{2},p\right)
-\frac{1}{9}g_{coh}\left(\bar{x}_{1},p\right)
g_{coh}\left(\bar{x}_{2},p\right)\bigr]
\cos(q\bar{x}_{12})
\Bigr\}\\
\\
=\int d^{4}\bar{x}_{1}d^{4}\bar{x}_{2}\,
\Bigl\{
g\left(\bar{x}_{1},p_1\right)g\left(\bar{x}_{2},p_2\right)
+\bigl(\bigl\langle | e_{i}e_{j}| ^{2}\bigr\rangle
-\frac{1}{9}(1+\delta _{ij})\bigr)
g_{coh}\left(\bar{x}_{1},p_1\right)
g_{coh}\left(\bar{x}_{2},p_2\right) \Bigr. \\
\\ \Bigl.\bigl.
+\delta _{ij}
g\left(\bar{x}_{1},p\right)g\left(\bar{x}_{2},p\right)
\cos(q\bar{x}_{12})
\Bigr\}\\
\\
=\int d^{4}\bar{x}_{1}d^{4}\bar{x}_{2}\,
\Bigl\{
g_{ch}\left(\bar{x}_{1},p_1\right)g_{ch}\left(\bar{x}_{2},p_2\right)+
\frac13\bigl(
g_{ch}\left(\bar{x}_{1},p_1\right)g_{coh}\left(\bar{x}_{2},p_2\right)+
g_{coh}\left(\bar{x}_{1},p_1\right)g_{ch}\left(\bar{x}_{2},p_2\right)
\bigr)
\Bigr. \\
\\ \Bigl.
+\bigl\langle | e_{i}e_{j}| ^{2}\bigr\rangle
g_{coh}\left(\bar{x}_{1},p_1\right)
g_{coh}\left(\bar{x}_{2},p_2\right)
+\delta _{ij}\bigl[
g_{ch}\left(\bar{x}_{1},p\right)g_{ch}\left(\bar{x}_{2},p\right)+
\frac23
g_{ch}\left(\bar{x}_{1},p\right)g_{coh}\left(\bar{x}_{2},p\right)
\bigr]
\cos(q\bar{x}_{12})
\Bigr\}
\label{a4n'}
\end{array}
\end{equation}
and recover Eqs.~(\ref{25}) for the pure QS correlation functions.

In the case of absent coherent emission, i.e. $d=g_{coh}=0$,
and on the usual assumption ($R_T^2\ll R_{ch}^2$)
of sufficiently smooth four--momentum
dependence of the chaotic emission function $g_{ch}(\bar{x},p)$
as compared with a sharp $q$--dependence of the QS and FSI correlations
(determined by the inverse characteristic distance between the
emission points), the chaotic emission functions in Eq.~(\ref{a4n})
can be taken out of the integral over $\kappa$ at small values of
$\kappa$, this integral thus being close to $\delta^4(\epsilon)$.
Choosing the momentum arguments in $g_{ch}$-functions in
accordance with Eq.~(\ref{a4n'}) for the case of absent FSI,
we get for the two--pion spectrum and the correlation function:
\begin{equation}
\begin{array}{c}
\omega _{{\bf p}_{1}}\omega _{{\bf p}_{2}}\frac{d^{6}N_{ij}}{d^{3}{\bf p}%
_{1}d^{3}{\bf p}_{2}}\approx
\int d^{4}\bar{x}_{1}d^{4}\bar{x}_{2}\cdot \\
\\
\left\{
g_{ch}\left(\bar{x}_{1},p_1\right)
g_{ch}\left(\bar{x}_{2},p_2\right)
|\psi_{q}^{(-)ij}(\bar{x}_{12})|^2+
\delta _{ij}
g_{ch}\left(\bar{x}_{1},p\right)
g_{ch}\left(\bar{x}_{2},p\right)
\psi_{q}^{(-)ij}(\bar{x}_{12})
\psi_{-q}^{(-)ij*}(\bar{x}_{12})\right\},
\label{a4'''}
\end{array}
\end{equation}
\begin{equation}
C_{ch}^{ij}\approx
\bigl\langle
|\psi_{q}^{(-)ij}(\bar{x}_{12})|^2
\bigr\rangle_{ch}
+
\delta _{ij}
\bigl\langle
\psi_{q}^{(-)ij}(\bar{x}_{12})
\psi_{-q}^{(-)ij*}(\bar{x}_{12})
\bigr\rangle_{ch}',
\label{cf1}
\end{equation}
where the average $\langle{\cal A}\rangle_{ch}$ and
quasi-average $\langle{\cal A}\rangle'_{ch}$ are defined as:
\begin{equation}
\bigl\langle {\cal A}\bigr\rangle_{ch}=
\frac{\int d^{4}\bar{x}_{1}d^{4}\bar{x}_{2}\, {\cal A}\,
g_{ch}\left(\bar{x}_{1},p_1\right) g_{ch}\left(\bar{x}_{2},p_2\right) }
{\int d^{4}\bar{x}_1\, g_{ch}\left(\bar{x}_1,p_1\right)
\int d^{4}\bar{x}_2\, g_{ch}\left(\bar{x}_2,p_2\right)},
\label{aver}
\end{equation}
\begin{equation}
\bigl\langle {\cal A}\bigr\rangle_{ch}'=
\frac{\int d^{4}\bar{x}_{1}d^{4}\bar{x}_{2}\, {\cal A}\,
g_{ch}\left(\bar{x}_{1},p\right) g_{ch}\left(\bar{x}_{2},p\right) }
{\int d^{4}\bar{x}_1\, g_{ch}\left(\bar{x}_1,p_1\right)
\int d^{4}\bar{x}_2\, g_{ch}\left(\bar{x}_2,p_2\right)}.
\label{aver'}
\end{equation}

In the case of a nonzero coherent contribution, the
$\epsilon/2$- and $\bar{x}_{12}$--dispersions in the pure
coherent term in Eq.~(\ref{a4n}) are the same ($2R_{coh}^2$),
contrary to usually negligible $\epsilon/2$-dispersion
in the pure chaotic term: $2R_{T}^2\ll 2R_{ch}^2$.
As for the mixed term, the $\epsilon/2$-dispersion would be
negligible if only the characteristic size $R_{coh}$ of the coherent
source were sufficiently small; with the increasing $R_{coh}$,
this dispersion may become important -
for $R_{coh}\approx R_{ch}$ it amounts to about half
of the $\bar{x}_{12}$--dispersion.
Therefore, the $\epsilon$--dependence of the Bethe--Salpeter
amplitudes should be generally retained in these terms.
The important exception is the case of practical interest in
heavy ion collisions, when the two charged pions are
created in their c.m.s. at a distance much larger than the
corresponding s--wave scattering length (of a fraction of fm)
and much smaller than their Bohr radius (of 387.5 fm).
The two--pion FSI interaction at small $q$ is then dominated by
the Coulomb FSI and depends only weakly on the space--time
separation of the emission points. In this case,
\begin{equation}
C_{ch}^{ij}\approx
\bigl\langle
|\psi_{q}^{(-)ij}(\bar{x}_{12})|^2
\bigr\rangle
+
\delta _{ij}
\bigl\langle
\psi_{q}^{(-)ij}(\bar{x}_{12})
\psi_{-q}^{(-)ij*}(\bar{x}_{12})
\bigr\rangle'+
\bigl(9\bigl\langle |e_{i}e_{j}|^{2}\bigr\rangle-1-\delta_{ij}\bigr)
G(p_{1})G(p_{2})\bigl\langle
|\psi_{q}^{(-)ij}(\bar{x}_{12})|^2
\bigr\rangle_{coh},
\label{cf1'}
\end{equation}
where the averages are defined as in Eqs.~(\ref{aver}) and (\ref{aver'})
with the substitutions $g_{ch}\rightarrow g$ or
$g_{ch}\rightarrow g_{coh}$ and,
the relative coherent contribution $G(p)$ - in
Eq.~(\ref{23}) with a formal substitution
$\int\limits_{\sigma _{out}}d\sigma _{\mu }p^{\mu}f(x,p)
\rightarrow \int d^{4}x g\left(x,p\right)$.

\section{Extracting coherent component of particle radiation}

Up to now, we have ignored the contributions $d^{3}N_{i}^{(l)}/d^{3}{\bf p}$
arising in the pion spectra from the decays of long--lived ($l$) sources
such as $\eta $-, $\eta ^{\prime }$--mesons, and also the unregistered
kaons and hyperons. The pions from these sources possess no
observable FSI (due to very large relative distance of the emission points)
as well as no noticeable interference effect, because the corresponding
correlation width is much smaller than the relative momentum resolution $%
q_{\min }$ of a detector. Therefore the measured correlation functions,
defined in Eq. (\ref{1}), can be expressed through the correlation functions
$\widetilde{C}^{ij}(p,q)$ (discussed in previous Section) of all pion pairs
$\pi ^{i}\pi ^{j}$ except for
those containing pions from long--lived sources
as follows \cite{LP79}:
\begin{equation}
C^{ij}(p,q)=n_{ij}(p_{1},p_{2})/n_{i}(p_{1})n_{j}(p_{2})=\Lambda ^{ij}(p)%
\widetilde{C}^{ij}(p,q)+1-\Lambda ^{ij}(p),  \label{C}
\end{equation}
where the suppression parameter $\Lambda ^{ij}(p)$ measures the fraction of
pion pairs containing no pions from long--lived sources:\footnote
{
One can include in $N_{i}^{(l)}$ and the corresponding suppression
parameters $\Lambda ^{ij}$ the contribution of misidentified particles which
also introduce practically no correlation.
}
\begin{equation}
\Lambda ^{ij}(p)=\left( 1-\frac{d^{3}N_{i}^{(l)}/d^{3}{\bf p}}{%
d^{3}N_{i}/d^{3}{\bf p}}\right) \left( 1-\frac{d^{3}N_{j}^{(l)}/d^{3}{\bf p}%
}{d^{3}N_{j}/d^{3}{\bf p}}\right) <1.  \label{14}
\end{equation}

In the (artificial) case of absent FSI effect, the correlation function
$\widetilde{C}^{ij}(p,q)=C_{QS}^{ij}(p,q)$,
and the averaging in $\langle\cos (qx_{12})\rangle' $
in the QS correlation functions in Eqs. (\ref{25}) should
be applied only to the pion pairs containing no pions from long--lived
sources. Then, assuming sufficiently good detector resolution, $q_{\min }\ll
R^{-1}$, we can determine the intercepts $C^{ij}(p,0)$ calculating the
correlation functions at $|q|\sim q_{\min }$:
\begin{equation}
C^{ij}(p,q_{\min }) =1+\Lambda ^{ij}(p)\Bigl[\delta_{ij}+
\bigl(9\bigl\langle |e_i e_j|^2\bigr\rangle
-1-\delta_{ij}\bigr)G^{2}(p)\Bigr].
\label{inter1}
\end{equation}
The intercepts are lower than 2 for any system of identical pions and they
are higher (lower) than 1 for $\pi ^{+}\pi ^{-}$ ($\pi ^{\pm }\pi ^{0}$)
systems.

Since the suppression parameters $\Lambda (p)$ are generally different for
different pion pairs, {e.g.}, due to different contributions of hyperon
decays, it is impossible, using only apparent intercepts in Eq.~(\ref{inter1}),
to separate the contributions of the coherent and long-lived sources, unless
there is known a ratio of the suppression parameters $\Lambda (p)$ for
identical and non-identical pions: $\Lambda ^{ii}(p)/\Lambda ^{ij}(p)$.
Then, for example, from the intercepts of the $\pi ^{+}\pi ^{+}$ and $\pi
^{+}\pi ^{-}$ correlation functions, one obtains the coherent fraction
squared:
\begin{equation}
G^{2}(p)=\frac{\Lambda ^{++}(p)}{\Lambda ^{+-}(p)}\left[ \frac{4}{5}\frac{%
\Lambda ^{++}(p)}{\Lambda ^{+-}(p)}+\frac{1}{5}\frac{C^{++}(p,q_{\min })-1}{%
C^{+-}(p,q_{\min })-1}\right] ^{-1}.  \label{20a}
\end{equation}

In fact, the knowledge of the ratio $\Lambda ^{ii}(p)/\Lambda ^{ij}(p)$ is
not of principle importance for the extraction of the coherent fraction $G(p)$.
Besides the intercepts, one
can exploit also the $q$ dependence of $C_{QS}(p,q)$ in
sufficiently wide interval to follow Eqs. (\ref{25}), and perform
simultaneous or separate fits of the correlation functions $C^{ij}$,
suitably parameterizing the correlator
$\langle \cos (qx_{12})\rangle $
and the function $G(p\pm {q}/{2})$.
For example, one can use the usual Gaussian correlator parameterization
\begin{equation}
\langle \cos (qx_{12})\rangle _{ch}^{\prime}\simeq \exp
(-q_x^2R_x^2-q_y^2R_y^2-q_z^2R_z^2)  \label{19.3}
\end{equation}
in the longitudinally comoving system (LCMS) in which the pion pair is
emitted transverse to the collision axis ($p_L=0$). The components of the
vector ${\bf q}$ are chosen parallel to the collision axis ({\rm z}%
=Longitudinal), parallel to the vector ${\bf p}_t$ ({\rm x}=Outward) and
perpendicular to the production plane ({\rm z,x}) of the pair ({\rm y}%
=Sideward). Assuming the same radii also for the coherent emission region,
and a transverse thermal law $\exp(-m_t/T)$ for the chaotic radiation with
the temperature $T$ ($m_t$ is the pion transverse mass), we can parameterize
the coherent fraction $G(p)$ similar to Eq. (\ref{rat-def}) for the
non-relativistic case with \cite{Akkelin}
\begin{equation}
D(p)\simeq D(0)\exp \left[-2(p_x^2R_x^2+p_y^2R_y^2+p_z^2R_z^2)+\frac{m_t}{T})%
\right],  \label{D-rare}
\end{equation}
and use Eq. (\ref{25e}) to calculate $\langle\cos(qx_{12})\rangle^{\prime}$.

The presence of the FSI effect introduces the additional $q$--dependence of
the correlation functions and thus improves, in principle, the accuracy of
the extraction of the coherent contribution $G(p)$. Consider, for example,
only effect of the Coulomb FSI and assume that the emission functions, $g_{ch}$
and $g_{coh}$, are localized in the regions of characteristic sizes much
smaller than the two--pion Bohr radius $|a|=387.5$ fm so that the modulus
of the non--symmetrized Coulomb wave function can be substituted by its value
at zero separation.
As a result the Coulomb effect factorizes in a form of so
called Gamow or Coulomb factor $A_{c}(ak^{\ast })=\left| \psi
_{q}^{coul}(0)\right| ^{2}$ (see, e.g., \cite{GKW}):
\begin{equation}
\widetilde{C}(p,q)=A_{c}(ak^{\ast })C_{QS}(p,q),~~~A_{c}(x)=(2\pi /x)/[\exp
(2\pi /x)-1],  \label{Ac}
\end{equation}
where $k^{\ast }=|{\bf q}^{\ast }|/2$ is momentum of one of the two pions in
their c.m.s.
For the correlation functions of like ($a=|a|$) and unlike ($a=-|a|$)
charged pions, we get
\begin{eqnarray}
C^{\pm\pm}(p,q) &=& \Lambda^{\pm\pm}(p)A_c(|a|k^*) \left[1+\langle\cos(qx_{12})%
\rangle^{\prime}-\frac45G(p+q/2)G(p-q/2)\right] +[1-\Lambda^{\pm\pm}(p)],
\nonumber \\
C^{+-}(p,q)&=& \Lambda^{+-}(p)A_c(-|a|k^*)\left[1+\frac15G(p+q/2)G(p-q/2)%
\right] +[1-\Lambda^{+-}(p)].  \label{17}
\end{eqnarray}
Similar to the case of absent FSI, we can again use the parameterizations (%
\ref{19.3}), (\ref{D-rare}) and the relation (\ref{25e}), and fit,
simultaneously or separately, the
correlation functions of like and unlike charged pions according to Eqs. (%
\ref{17}). Moreover, the known $q$--dependence of the Gamow factors allows
to separate the coherent fraction $G(p)$ from the suppression parameter $%
\Lambda(p)$ in a model independent way, without exploiting the
q--dependence of $\langle\cos(qx_{12})\rangle_{ch}$ and $G(p\pm q/2)$.
Indeed, one can perform the fits according to
Eqs. (\ref{17}) in an interval of $q_{\min}<|q|\ll R^{-1}$ guaranteeing $%
\langle\cos(qx)\rangle^{\prime}\approx 1$ and $G(p_{1,2})\approx G(p)$.
The q--dependence of the correlation functions is then uniquely
determined by the known functions $A_{c}(|a|k^{\ast })$ and $%
A_{c}(-|a|k^{\ast })$, and the three fitted parameters: $G({p}%
)$, $\Lambda ^{\pm \pm }(p)$ and $\Lambda ^{+-}(p)$.
Of course, such an analysis requires very good detector
resolution and its good understanding.

Note that Eqs. (\ref{17}) are not applicable for very small
($\sim 1$ fm) as well as for large sources. In the former case one has to
account for the strong FSI, in the latter - for the finite--size Coulomb
effects. For ultra-relativistic heavy ion collisions, the strong FSI
effect on two--pion correlation functions is
negligible for like charge pions and small (a few percent) for unlike pions.
The Coulomb finite--size effects can be approximately taken into account,
substituting the Gamow factor $A_{c}(ak^{\ast })$ in Eqs. (17)
by the finite size Coulomb factor $\widetilde{A}_{c}(ak^{\ast
},\langle r^{\ast }\rangle /a)$ \cite{Sin}.
The latter represents a simple function of the arguments $ak^{\ast }$ and
$\langle r^{\ast }\rangle /a$,
where $\langle r^{\ast }\rangle $ is the mean distance of the
pion emission
points in the pair c.m.s., corresponding to a given momentum ${\bf p}$.
Particularly, $\widetilde{A}_{c}\doteq A_{c}(ak^{\ast })
[1+2\langle r^{\ast
}\rangle /a]$ at $k^{\ast }<\sim
1/\langle r^{\ast }\rangle $.

The dependence of the Coulomb factor on the unknown parameter $\bigl\langle
r^{\ast }\bigr\rangle $ somewhat complicates the model-independent
method for the
extraction of coherent component $G(p)$ exploiting only the correlation
functions in the region of very small relative momenta.
Now, the simultaneous analysis of the correlation
functions of like and unlike charged pions is required because their
separate analysis yields the coherent contribution $G(p)$ up to a correction
$\langle r^{\ast }\rangle /a$ only. As for the method based on
a fit in a wide $|{\bf q}|$--interval, the quantity
$\langle r^{\ast }\rangle $ being
a unique function of the parameters characterizing the emission density,
actually represents no new free parameter. Particularly,
for a universal anisotropic Gaussian ${\bf r}^{\ast }$%
--distribution of the chaotic and coherent emission functions,
the quantity $\langle r^{\ast}\rangle $
can be expressed analytically through the Gaussian interferometry
radii $R_{y}$, $R_{z}$ and $R_{x}^{\ast }=\frac{M_{t}}{M}R_{x}$ ($M$ and $%
M_{t}$ are the two--pion effective and transverse masses respectively) in
the case of practical interest, when $R_{x}^{\ast }\geq R_{y}\approx R_{z}$
\cite{Sin}.

In practice, however, the Gaussian parametrization of the relative
distances between the emission points
may happen to be insufficient. Particularly, it can
lead to apparent inconsistencies in the treatment of QS and FSI effects
because the latter is more sensitive to the tail of the distribution of the
relative distances.
If, for example, the $r^*$--distribution were represented by a
sum of two Gaussians with essentially different
mean squared radii, the $r^*$--''tail'',
determined by the larger Gaussian radius, would influence
the observed correlation
functions in different ways. For identical
pions, the ''tail'' results in an additional rather narrow peak in the QS
correlation function; however, this ''tail'' would show up only as
a suppression of the
correlation function if the peak were concentrated at $q\lesssim q_{\min }$ or
if one measured a given projection of the correlation function (e.g., in $%
q_{side}$ direction) fixing others ($q_{long}$ and $q_{out}$) in
the interval exceeding the width of the narrow peak.
At the same time, the $r^*$--''tail'' would influence Coulomb correlations at small
$q\gtrsim $ $q_{\min }$
since the long-distance nature of Coulomb forces leads to the observable
effect conditioned by the ''tail'' up to $r^* \sim |a|$.
In such a situation, one can no more rely on the equality between $%
\langle r^{\ast }\rangle _{QS}$, determined by the interferometry radii, and
the characteristic size $\langle r^{\ast }\rangle _{C}$ determining the
Coulomb FSI effect. Generally, one has to introduce also different
suppression parameters $\Lambda _{QS}<\Lambda _{C}$ corresponding to $%
\langle r^{\ast }\rangle _{QS}<\langle r^{\ast }\rangle _{C}$. Eqs. (\ref{17}%
) for the correlation functions of like and unlike charged pions, with the
substitution of the Gamow factor $A_{c}(ak^{\ast })$ by the finite--size
Coulomb factor $\widetilde{A}_{c}(ak^{\ast },\langle r^{\ast }\rangle /a)$
\cite{Sin}, are then modified to the form:
\begin{eqnarray}
C^{\pm \pm }(p,q) &=&\Lambda _{QS}^{\pm \pm }(p)\widetilde{A}_{c}(|a|k^{\ast
},\langle r^{\ast }\rangle _{QS}^{\pm \pm }/|a|)\left[ \langle \cos
(qx)\rangle ^{\prime }-\frac{4}{5}G(p+q/2)G(p-q/2)\right] +  \nonumber \\
&&\Lambda _{C}^{\pm \pm }(p)\widetilde{A}_{c}(|a|k^{\ast },\langle r^{\ast
}\rangle _{C}^{\pm \pm }/|a|)+[1-\Lambda _{C}^{\pm \pm }(p)],
\nonumber \\  \label{17a} \\
C^{+-}(p,q) &=&\Lambda _{QS}^{+-}(p)\widetilde{A}_{c}(-|a|k^{\ast },-\langle
r^{\ast }\rangle _{QS}^{+-}/|a|)\frac{1}{5}G(p+q/2)G(p-q/2)+  \nonumber \\
&&\Lambda _{C}^{+-}(p)\widetilde{A}_{c}(-|a|k^{\ast },-\langle r^{\ast
}\rangle _{C}^{+-}/|a|)+[1-\Lambda _{C}^{+-}(p)]. \nonumber
\end{eqnarray}
To simplify the analysis, one can neglect a small difference between the
suppression parameters $\Lambda _{QS}$ and $\Lambda _{C}$ due to the tail of
the $r^{\ast }$--distribution and also neglect a presumably small difference
between $\langle r^{\ast }\rangle ^{\pm \pm }$ and $\langle r^{\ast }\rangle
^{+-}$.

Note, that at SPS and RHIC energies the effect of strong FSI on $\pi ^{+}\pi ^{-}$
correlations is still quite noticeable and, when neglected, it can lead to a
suppression of a fitted $\langle r^{\ast }\rangle ^{+-}$
by $\sim 50\%$.
Also, due to a substantial inaccuracy of the
Coulomb factor $\widetilde{A}_{c}(ak^{\ast },
\langle r^{\ast }\rangle /a)$
near the tailing point $k^{\ast }\sim
1/\langle r^{\ast }\rangle ,$ the
parameters $\langle r^{\ast }\rangle ^{++}$ and
$\langle r^{\ast }\rangle^{+-}$ can be respectively overestimated
and underestimated if the fitted region
were not sufficiently wide.
Further, in the case of different chaotic and coherent emission volumes,
one has to use finite--size Coulomb factors with
different $\langle r^{\ast }\rangle $ in the chaotic, coherent
and mixed terms.
All these problems can be overcome exploiting the exact formulae
for the two--pion wave functions (in the equal time approximation)
and calculating the correlation
functions according to the approximate Eq.~(\ref{cf1'}).
To control the systematic errors due to the smoothness
assumption in Eq.~(\ref{cf1'}), one can give up this assumption
(at least in the pure coherent term) and check the results
using instead the general expression for the two--pion spectrum
in Eq.~(\ref{a4n}).

After the extraction of the fractions $G(p)$ and $\Lambda ^{++}(p)$
or $\Lambda ^{--}(p)$, one can obtain the coherent part of the measured
single--pion spectra $\omega _{{\bf p}}d^{3}N_{\pm }/d^{3}{\bf p}$. Using
Eq. (\ref{14}), and substituting $d^{3}N/d^{3}{\bf p\rightarrow (}%
d^{3}N_{\pm }/d^{3}{\bf p-}d^{3}N_{\pm }^{(l)}/d^{3}{\bf p)}$ in Eq. (\ref
{23}), one gets:
\begin{equation}
\omega _{{\bf p}}\frac{d^{3}N_{coh}}{d^{3}{\bf p}}\equiv \frac{1}{3}\left|
d({p})\right| ^{2}=\omega _{{\bf p}}\frac{d^{3}N_{\pm }}{d^{3}{\bf p}}%
G(p)\sqrt{\Lambda ^{\pm \pm }(p)}.  \label{22}
\end{equation}
The coherent part of the observed spectra is thus directly
connected with the intensity $\left| d({p})\right| ^{2}$ of
the quasi-classical source of coherent pions.

\section{Conclusions}

Using the density matrix formalism, satisfying the requirements of the
isospin symmetry and the super-selection rule for generalized coherent
states, and accounting for the final state interaction in the two--body
approximation, we have developed methods allowing one to study the coherent
component of pion radiation which, in heavy ion collisions, is likely
conditioned by formation of a quasi-classical pion source.

These methods are based on a nontrivial modification of the effects of
quantum statistics and final state interaction on two--pion correlation
functions (including those of non-identical pions) in the presence of a
coherent pion radiation generated by the decay of the quasipionic ground
state (''condensate''). It has been shown that the combined analysis of the
correlation functions of like and unlike pions gives the possibility to
discriminate between the suppression of the like--pion
correlation functions conditioned by the coherent pion component
and that due to the decays of long--lived sources.

The methods allowing to extract the coherent pion component from $\pi
^{+}\pi ^{-}$ and $\pi ^{\pm }\pi ^{\pm }$ correlation functions and
single--pion spectra have been discussed in detail for large expanding
systems produced in ultra--relativistic heavy ion collisions. For such
systems, the two--pion final state interaction is dominated by the Coulomb
one and plays an important role in this analysis,
allowing one to determine the coherent fraction using a
suitable model for the coherent and chaotic
emission functions and fitting the corresponding correlation functions. For
rough estimations the procedure can be substantially simplified accounting
for the finite--size Coulomb effects in an approximate analytic form \cite
{Sin}.

Finally, the coherent fractions extracted from the correlation analysis,
combined with the single--pion spectra, give us the possibility to determine
the spectrum of the coherent pion radiation above the thermal background
and,
therefore, to estimate the quasipionic condensate at the pre-decaying stage
of the matter evolution and discriminate
between possible mechanisms of coherent production in ultra--relativistic
A+A collisions.

\section{Acknowledgments}

This work was supported by French-Ukrainian grant No. Project 8917, by
Ukrainian - Hungarian Grant No. 2M/125-99, by Ukrainian-German Grant No.
2M/141-2000, and by GA Czech Republic, Grant No. 202/01/0779. We gratefully
acknowledge Barbara Erazmus and Edward Sarkisyan
for the interest in this work and fruitful
discussions.

\end{document}